\documentclass[10pt,twocolumn,notitlepage]{article}
\usepackage{graphicx}
\usepackage{latexsym}
\newcommand{\figurewidth}{8cm}       %for twocolumn

\begin{document}
\title{On orientational relief of inter-molecular potential\\ and
the structure of domain walls in fullerite C$_{60}$}
\author{Julia M.~Khalack,$^{1,2,\dag}$
Vadim M.~Loktev$^{1,\ddag}$
\\ \small \it
%} \affiliation{
$^1$Bogolyubov Institute for Theoretical Physics
of the National Academy of Sciences of Ukraine, \\ \small
\it
14b Metrologichna Str., Kyiv-143, 03143 Ukraine \\ \small
\it
$^2$Stockholm University, Arrhenius Laboratory, 
Division of Physical Chemistry,
S-106 91 Stockholm, Sweden \\ \small
$^\dag$E-mail:
%julia@nonlin.bitp.kiev.ua,
julia@physc.su.se \\ \small
$^\ddag$E-mail:
vloktev@bitp.kiev.ua
}
\maketitle
\sloppy
\begin{abstract}
\em
\small
A simple planar model
for an orientational ordering of threefold
molecules on a triangular lattice modelling
a close-packed (111) plane of fullerite
is considered.
The system has 3-sublattice ordered ground state
which includes 3 different molecular orientations.
There exist 6 kinds of orientational domains,
which are related with a permutation or a mirror
symmetry. Interdomain walls are found to be
rather narrow.

The model molecules have two-well orientational
potential profiles, which are slightly
effected by a presence of a straight domain
wall. The reason is a stronger correlation between
neighbour molecules in triangular lattice
{\em versus} previously considered square lattice

A considerable reduction (up to one order)
of orientational interwell potential barrier
is found in the core regions of essentially
two-dimentional potential defects,
such as a three-domain boundary or a kink in the domain wall.
For ultimately uncorrelated nearest neighbours
the height of the interwell
barrier can be reduced even by a factor of 10$^2$.

\end{abstract}
\maketitle

\section{Introduction}

An elegant hollow cage structure of the C$_{60}$ fullerene molecule has
drawn a close attention of scientists because of its unique $I_h$
icosahedron symmetry. A nearly spherical form of the molecule
leads to very unusual physical properties of solid C$_{60}$,
fullerite.\cite{L92,R94,G97,KBK00}
While at the room temperature the molecules can be considered
to be the exact spheres, the low temperature properties of fullerite are
determined by the deviation of the molecule geometry from the
spherical one. At these temperatures an orientational molecule
ordering takes place, which is a basic issue for understanding the
results of recent He-temperature experiments on
heat transport,\cite{EMDN97,EMD99}
linear thermal expansion,\cite{AEMSSU97,AGEMSUM01}
and the specific heat \cite{OTP93} of the C$_{60}$
fullerite.

The mass centers of the C$_{60}$ molecules in fullerite
are arranged into an \emph{fcc} structure characteristic
for close packed spheres with isotropic interactions between them.
At the room temperature the molecules are found to be freely
rotating. The resulting crystal space group is $Fm\bar{3}m$.

Upon the lowering of temperature,
the fullerite is subjected to two transitions.
At $T \approx 260$~K,
it undergoes the first order phase transition, after which
the \emph{fcc} crystal lattice is divided into four
simple cubic sublattices.
The molecules are now allowed to rotate about one of the 10
molecular threefold axes. Other two of the three rotational
degrees of freedom are frozen.
Within each of the four sublattices,
the allowed molecule rotation axis is fixed along one of the four
($[111]$, $[1\bar{1}\bar{1}]$, $[\bar{1}\bar{1}1]$, or
$[\bar{1}1\bar{1}]$) threefold cubic axes, so that the
crystal space group is $Pa\bar{3}$.

It is worth noting that the truncated
icosahedron form of a C$_{60}$ molecule
allows for a more symmetric regular crystal structure
with only one sublattice and with the four of the above mentioned
molecule threefold axes oriented along the threefold crystal axes
(usually regarded as a standard molecule orientation).
But such a structure is energetically unfavourable for the
anisotropic intermolecular interaction.
Instead, the observed low-energy structure is obtained by
a simultaneous counterclockwise
22$^{\circ}$ rotation of the C$_{60}$ molecules from the initial
standard orientation about their fixed $Pa\bar{3}$ threefold axes.

As a result of such a rotation,
each C$_{60}$ molecule is oriented with one of
the negative charged double C=C bonds to
every one of the six neighbour molecules belonging
to the same close-packed (111) plane perpendicular to the
molecular rotation axis. To the other six neighbours (belonging to
two adjacent (111) planes) the molecule is oriented with the
positive charged pentagons (P). Therefore following a commonly
used notation we denote this
orientation as ``P orientation''. For an  ideal structure with
all the molecules having a P orientation, every pair of nearest
neighbours is characterized with a pentagon from one molecule
opposing a double bond from another molecule.

On the other hand, a potential profile of the fullerene
molecule rotating about its fixed threefold axis has an additional
metastable minimum%
\footnote{We do not consider to be distinct the energy degenerate
minima obtained by 120$^\circ$~rotation
about the threefold molecular axis.}
corresponding to 82$^\circ$ rotation from the
standard orientation (and to 60$^\circ$ rotation from the P
orientation). In this minimum, the molecule opposes the neighbour
molecules from the same (111)~plane with the double bonds,
and the molecules from adjacent planes are opposed with hexagons
(H orientation\footnote{%
Strictly speaking, the term `H (or P) configuration' is more adequate 
for describing a mutual orientation of two neighbouring molecules.
Nevertheless, for every chosen pair of neighbouring molecules 
(let us denote them as A and B) with the fixed directions 
of allowed rotation axes, the mutual orientation depends strongly 
only on the rotation angle of one molecule (say, A). 
The other molecule (B) is always (at any angle of its rotation) 
turned to the first one (A) with a double bond.
Therefore, the interaction energy of the pair weakly depends 
on the rotation angle of the second molecule.
As to the molecule A, it is at any rotations always turned to B 
with a belt of pentagons and hexagons interconnected by single bonds.
Thus, namely the molecule A of the pair A,B is responsible 
for the mutual orientation.
Aside from this, upon the rotation of the molecule A from 
orientation P to orientation H this molecule becomes turned with 
hexagons (instead of pentagons) to five more its nearest neighbours.
At the same time, the energy of its interaction with the other 6 nearest 
neighbours remains practically unchanged, because the energy depends 
mainly on the orientation of that latter molecules. 
Basing on the reasons mentioned above, we follow the common notations
and use the letters `P' and `H'
to denote an orientation of a single molecule, while keeping in mind
those 6 pair orientations for which this molecule rotation angle is crucial.
}).
The energy difference between the P and H minima is about 11~meV
($\approx$130~K)
and the height of the potential barrier is 235-280~meV
($\approx$2700-3200~K).\cite{YTSLM92,SHY97}

At the high enough temperatures, molecules are able to jump
between the two energy minima due to the processes of a thermal
activation. An average P/H ratio is given by the Boltzmann
distribution law. Just below the high temperature phase
transition (T$\approx$260~K) a fraction of the H~oriented
molecules is close to 0.5, and for T$\approx$90~K it tends to
0.15.\cite{DIDP92}

For the temperatures below 90~K the situation changes drastically.
A waiting time for a molecule to obtain a sufficient energy for a
jump between the P and H orientations reaches the order of several
days ($10^4-10^5$~s) or even more. Therefore at some critical
temperature (its exact value near 90~K depends slightly on the
cooling conditions)
the molecules become frozen in their current orientational
minima, and a transition to an orientational glassy phase
takes place. Below this transition a fraction of H oriented
molecules remains practically unchanged and equal to its equilibrium
value (about 15\%) characteristic
 for the temperature of the glass transition.
In other words, in the average every 7th molecule has the H
orientation, and with the very high probability every C$_{60}$
molecule has at least one misoriented neighbour.

While the orientational glass structure is believed to persist down
to the lowest temperatures, some of the experimental data obtained
at helium temperatures can not be explained with the concept of
H~oriented molecules alone.
For example, the data on heat conductivity \cite{EMDN97,EMD99}
show a maximum phonon mean free path of about 50 intermolecular
spacings, what implies only a 0.02 fraction of scattering
(``wrong'') molecules.
Besides, the negative thermal expansion \cite{AEMSSU97,AGEMSUM01}
and the linear contribution to the specific heat \cite{OTP93,TSB98}
of the fullerite samples at helium temperatures are
explained with the help of the tunnelling
(i.e. quantum) transitions of the
C$_{60}$ molecules between nearly degenerate orientational minima.
Such a possibility was firstly assumed in
Ref.~\cite{IL93}, where all the molecules in a crystal were
supposed to be in tunnelling states.
However, the paper \cite{TSB98}
accurately estimates the tunnelling frequency to be about
5.5~K, and the number of tunnelling degrees of freedom to be
$\sim 4.8\times10^{-4} (N/60)$, where $N$ is the number of carbon
atoms in a crystal.
Obviously, the number of the H~oriented molecules is much bigger,
and the above mentioned potential barrier between the H and P
orientations is too
high to provide such a low tunnelling frequency.
Therefore the defect states other than the simple H oriented
molecules should  be considered.

One of the possibilities for a realization of
the low potential barrier for C$_{60}$ molecule
is indicated in our previous paper.\cite{LPK01}
Relatively low barrier sites can appear within the orientational
domain walls because of the superposition of the mutually
compensating potential curves due to interaction with the neighbour
molecules belonging to different domains.
For the case of orientational ordering of hexagons on a square
planar lattice considered in \cite{LPK01},
the height of the potential barrier in the wall is found to be 5
times less than in the regularly ordered lattice.
Such a lowering seems to be insufficient to provide the necessary
magnitude of tunnelling frequency following from the available
experimental data analysis.\cite{TSB98}

Meanwhile, most of the results obtained for a square lattice seem
to be caused by the incompatibility of
the molecule threefold $C_3$ symmetry axis
with the lattice fourfold $C_4$ symmetry axis.
In the case of fullerite, a fullerene molecule holds
4 threefold axes and 3 twofold axes intrinsic to the \emph{fcc}
lattice. Furthermore, the closest-packed (111) plane of the
$Pa\bar{3}$ lattice has a hexagonal structure. Six of the 12
molecule nearest neighbours belong to a hexagonal plane,
while only 4 of them belong to the same square (001) plane.

Therefore it is interesting and necessary to investigate the main
features of orientational ordering for the case of the molecule
symmetry identical to that of the lattice.
In the present paper we are concerned with the
possible orientational domain structures formed by the
simple flat hexagon-shaped molecules
arranged into a more relevant to a real situation hexagonal
lattice, with both the molecule and the lattice symmetry axes
being $C_3$.
The main purpose of the paper is to estimate the energetic
inter-molecular interaction barriers
for both regular close-packed planar structure and
for the vicinity of extended orientational defects.

It is a pleasure and a honour for us to devote this paper to the
Ukrainian low temperature experimentalist of a world-wide
reputation Prof. Vadim~G. Manzhelii whose contribution to the
physics of cryocrystals in general and to the physics of
fullerites and fullerides in particular is well-known and can not
be overestimated.

\section{Model}

Let us consider a system of flat hexagonal molecules%
\footnote{In some sense, they can be regarded as
imitating the C$_{60}$ molecules viewed along C$_{3}$ axis.
Strictly speaking, such imitation is competitive
only for the fullerene
molecules with the fixed C$_{3}$ axis perpendicular to the
considered (111) plane. The C$_{60}$ molecules belonging
to other three
$Pa\bar{3}$ sublattices have their fixed threefold axes
tilted to this plane.
}
located in the sites of a rigid hexagonal planar lattice,
modelling a (111) plane  of the 3D \emph{fcc} lattice.

\begin{figure}[th]
  \centering
   \includegraphics[width=\figurewidth]{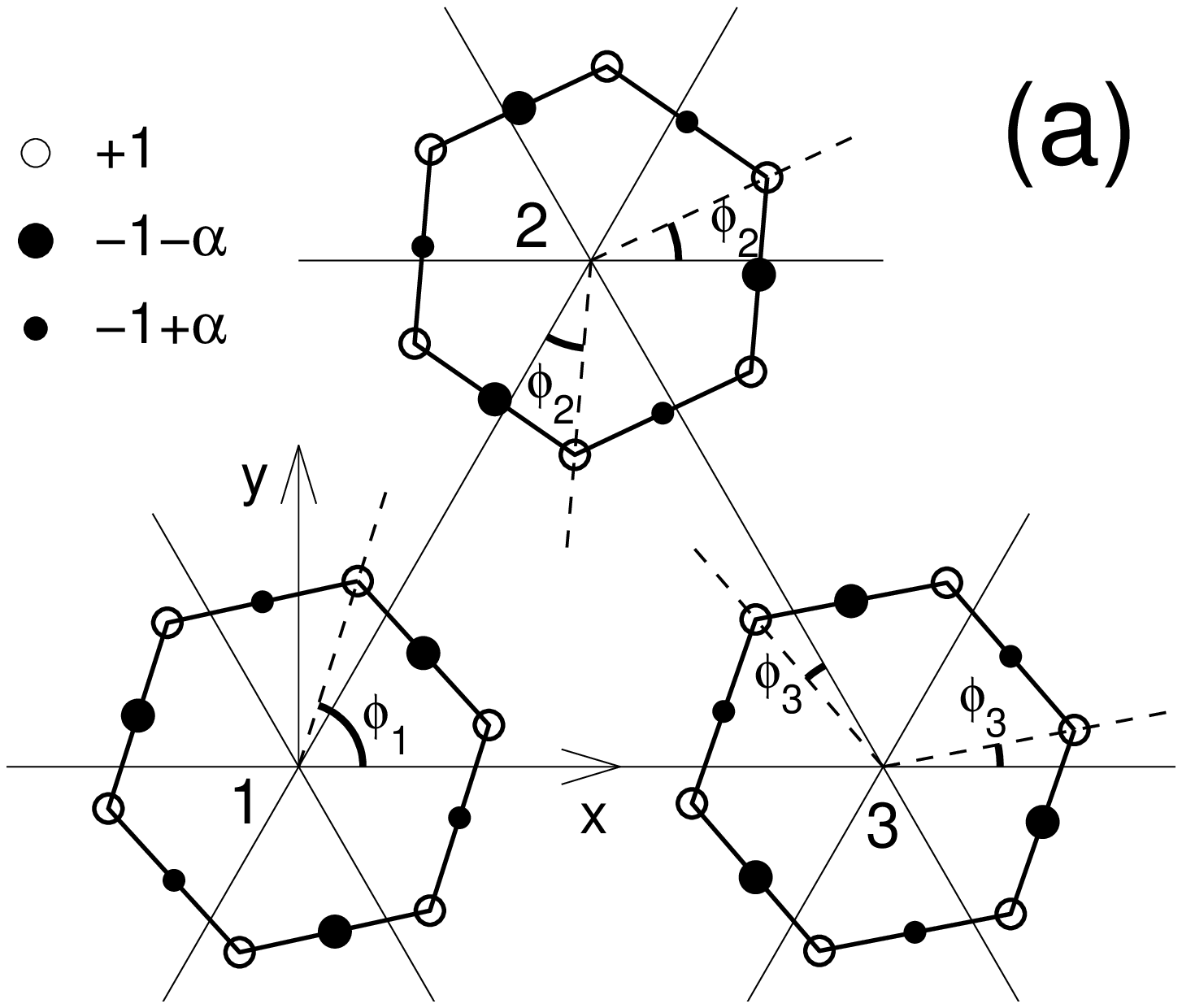}
   %\hfill
   \includegraphics[width=\figurewidth]{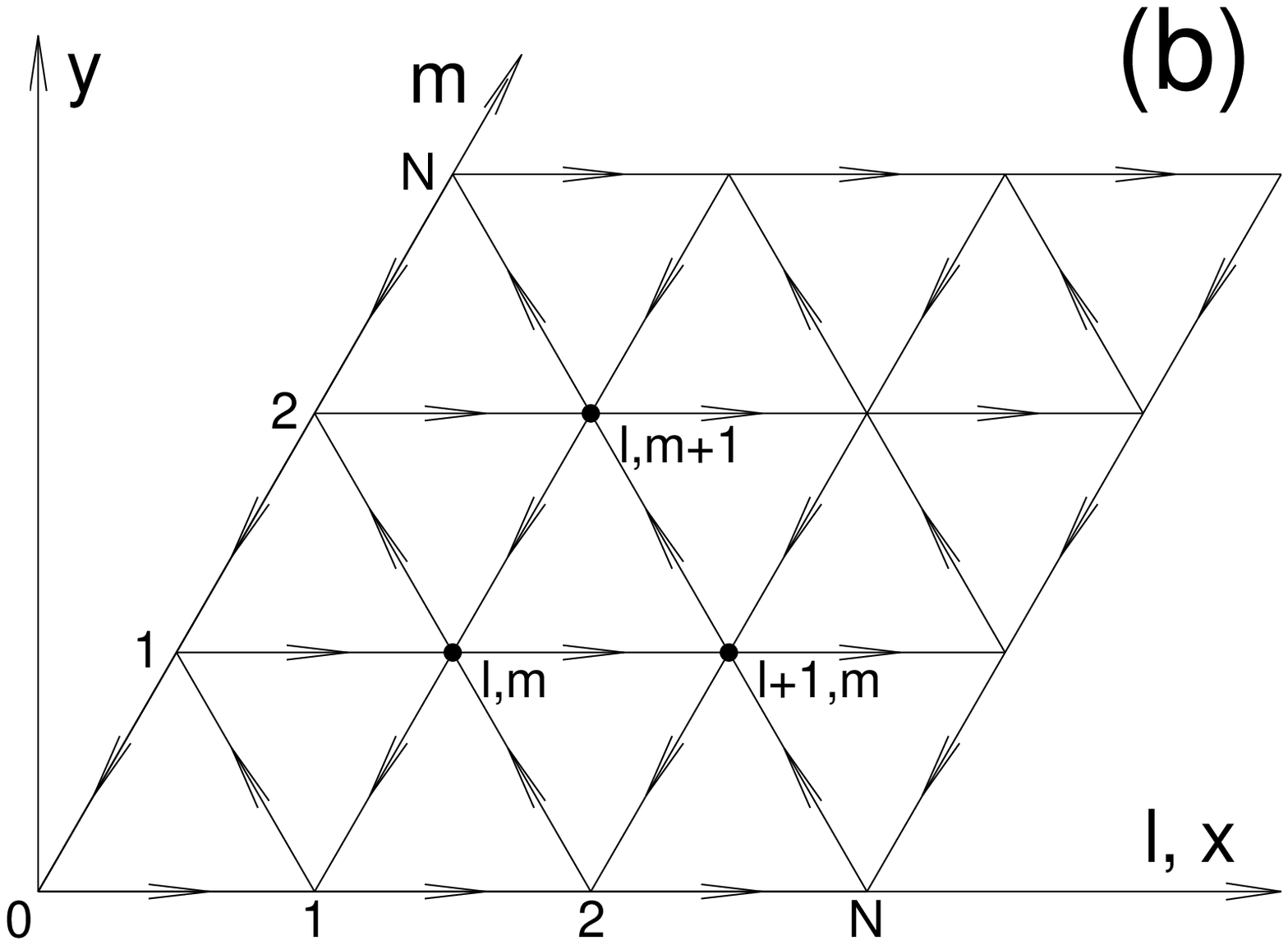}
   \caption{\label{fig:model}
   (a)~A local geometry of the model molecules on the triangular
   lattice. Note that molecular rotation angles (shown with the
   help of dashed lines)  can be measured
   from any of the three lattice directions.
   (b)~A geometry of the simulation cell.
   Arrows give the $1\rightarrow 2$ order of the input parameters
   for the pair interaction function $W(\phi_{1};\phi_{2})$.
   }
   \end{figure}

Following the empirical potentials used for modelling the
intermolecular fullerene interaction (see, for example,
Ref.~\cite{LRM97} and references therein),
we suppose two kinds of negative charges, $-(1\pm\alpha)$,
to be located at the centers of hexagon sides
(see large and small
filled circles in Fig.~\ref{fig:model}(a)).
These negative charges recall single and double covalent bonds
between carbon atoms at the hexagon edge of the truncated
icosahedral fullerene molecule.
Introduction of the charge parameter $\alpha$
reduces the $C_{6}$ hexagon symmetry down to $C_{3}$
intrinsic for real C$_{60}$.
A requirement of electro-neutrality of the model hexagon molecule
stipulates a presence of unit positive charges at its
vertices (shown with the open circles in Fig.~\ref{fig:model}(a)).

For an initial orientation (an analogue of the standard
orientation in fullerite) the molecules are chosen to be aligned
with the positive charges along the lattice directions.
The topmost (positive $Y$ direction)
negative charge has to be a larger one (see Fig.~\ref{fig:model}(a)).
The molecule rotation angle $\phi$ is measured starting from the
positive $X$ direction.

Interaction between the two nearest molecules is
given with the superposition of the Coulomb interactions
between the vertices and bonds of these molecules.
The exact form of the resulting interaction function
can be found in Ref.~\cite{LPK01} (Eqs.(1-4)).
Interaction is multipolar, so it depends not on the
difference of the molecules rotation angles only
(as for the case of the spin systems
with Heisenberg exchange coupling),
but essentially on both the angles.
So, the energy of interaction of the two neighbour molecules
characterized with rotation angles $\phi_{1}$ and $\phi_{2}$
has to be written as $W(\phi_{1};\phi_{2})$.
The clockwise and counterclockwise rotations have different effect
on the interaction:
\begin{equation}\label{eq1}
W(\phi_{1};\phi_{2})\ne
W(-\phi_{1};\phi_{2})\ne
W(\phi_{1};-\phi_{2}).
\end{equation}
On the other hand, a clockwise rotation of the first molecule is
somewhat equivalent to a counterclockwise rotation of the second
molecule. Hence, a combination of the lattice mirror symmetry with
the molecule mirror symmetry leads to the following symmetry
relation for the interaction function:
\begin{equation}\label{eq:mirror}
W(\phi_{1};\phi_{2})=
W(-\phi_{2};-\phi_{1}).
\end{equation}

Rotating the molecules shown in Fig.~\ref{fig:model}(a)
by an angle $2\pi/3$ (or $4\pi/3$)
about a threefold axis located in
the center of the triangle 123, one can
find that the pair interaction of the molecules 2,3 (or 3,1)
is given with the same function
$W(\phi_{2};\phi_{3})$ (or $W(\phi_{3};\phi_{1})$, respectively).
A relative displacement (which was
vertical or horizontal in the case of the
square lattice considered in Ref.~\cite{LPK01}) of the two
molecules does not have to be taken into account,
but an order of the angle parameters is essential.

For a simulation of the possible domain structures,
we consider a finite parallelepiped-shaped system,
which geometry is shown in Fig.~\ref{fig:model}(b).
It consists of $20\times 20$ hexagon molecules
labeled with the two indexes $l$ and $m$.
Arrows show the order of the interaction function arguments
for each pair of hexagons.
The system Hamiltonian then reads:
%\begin{widetext}
\begin{eqnarray}\label{Ham}\nonumber
  H&=&\sum_{l,m=0}^{N-1}
  \left[
  W(\phi_{lm};\phi_{l+1,m})\right.
  \\ \nonumber &+&\left.
  W(\phi_{l+1,m};\phi_{l,m+1})+
  W(\phi_{l,m+1};\phi_{lm})
  \right]
  \\ \nonumber &+&
  \sum_{l=0}^{N-1}
  W(\phi_{lN};\phi_{l+1,N})
  \\ &+&
  \sum_{m=0}^{N-1}
  W(\phi_{N,m+1};\phi_{Nm}),
\end{eqnarray}
%\end{widetext}
where $N=19$, and the last two terms are introduced to take into
account the edge molecules.
For numerical simulations, the charge parameter $\alpha$
is chosen to be 1/3.
A hexagon side makes 0.3 of the lattice spacing.

\section{Possible ordering types}

For a general case of an orientational ordering of the identical
molecules on a planar hexagonal lattice, the structures with
1, 3, 4, and 7 sublattices are possible.
One sublattice structure would correspond to a uniform rotation
of all the molecules on a lattice.
Three sublattice structure is characteristic of antiferromagnetic
systems (Loktev structure \cite{L_af}).
A close-packed (111) plane of the $Pa\bar{3}$ structure
should contain the
molecules belonging to four different sublattices.
The results of STM imaging of the (111) fullerite surface
\cite{WZWHLY01}
confirm this fact.\footnote{%
At the beginning of the fullerene era, there were some
publications \cite{TAVLM92,GPMSWM92} reporting an
8-sublattice {\em fcc} structure for the low temperature fullerite.
This structure could be obtained by division of each of the four
{\em sc} $Pa\bar{3}$ sublattices into two {\em fcc} sublattices
with different (P and H) molecular orientations.
However, the 8-sublattice structure has not been confirmed by
further investigations. Therefore, we do not consider it here.}
A more complicated case of seven sublattices could be expected for
less symmetric molecules.

\begin{figure}[th]
  \centering
   \includegraphics[width=\figurewidth]{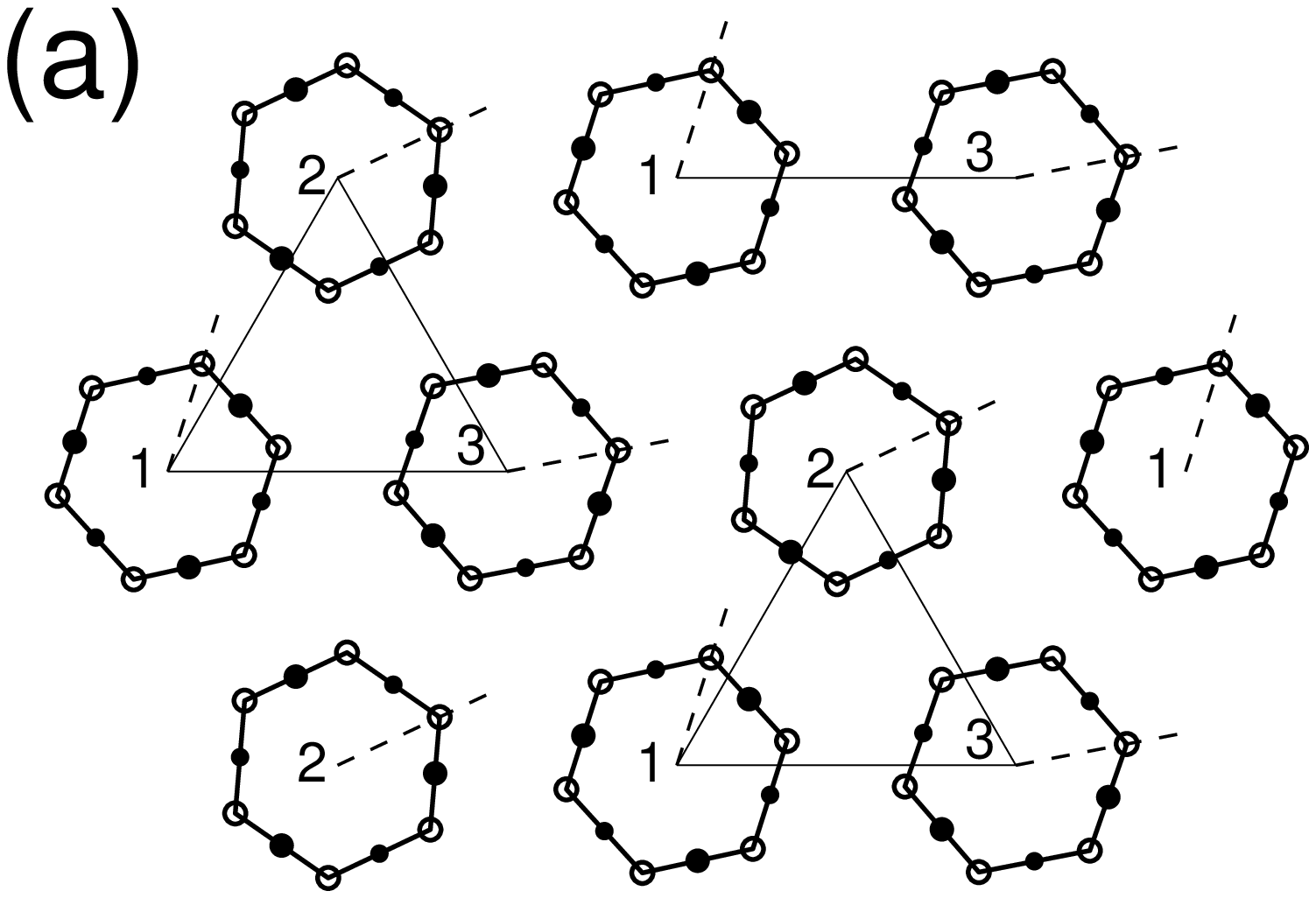}
   %\hfill
   \includegraphics[width=\figurewidth]{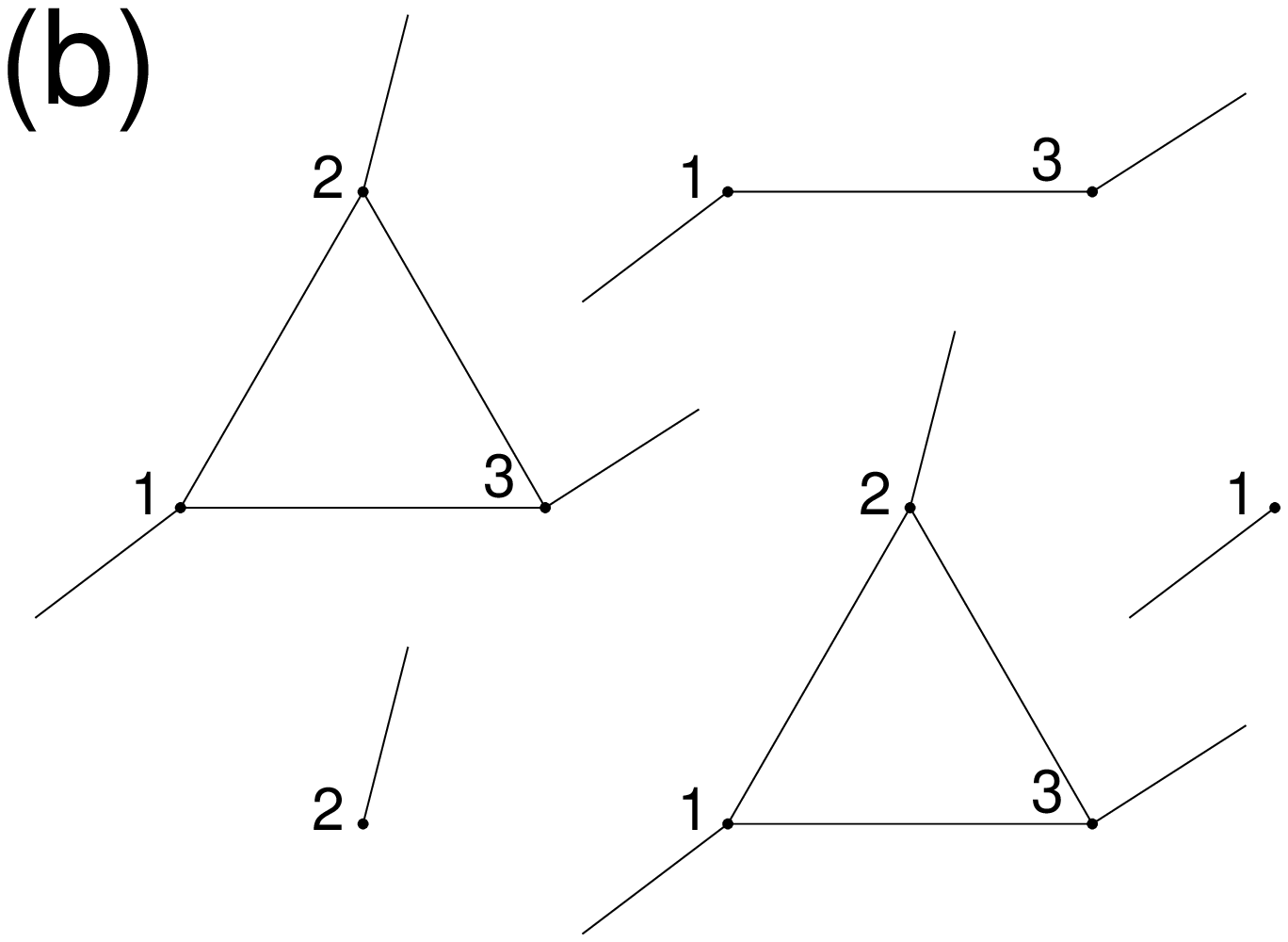}
   \caption{\label{fig:patterns}
   A ground state orientational ordering of the hexagon
   molecules~(a), and
   the same ordering patterned with vectors~(b).}
   \end{figure}

As to considered $C_3$ symmetric hexagons,
a rather aesthetic expectation of the threefold site symmetry%
\footnote{
An absence of the site symmetry would induce a distortion of the
lattice.}
implies either 1 or 3 sublattice case, or a 4 sublattice structure
involving three identical rotations.
Numerical calculations give for a ground state a three sublattice
structure shown in Fig.~\ref{fig:patterns}(a).
It is interesting to notice that in the 3 sublattice
structure the energy minimum corresponds to the molecule
positions which do nat provide the minimum
of the pair potential.
The obtained molecule rotation angles are ($\alpha=1/3$):
\begin{equation}\label{eq:angles}
 \phi_1=72.37209^\circ ; \;
 \phi_2=25.24477^\circ ; \;
 \phi_3=10.87533^\circ .
\end{equation}
Another possible (energy degenerate) ground state can be found
with the help of the symmetry relation (\ref{eq:mirror}).
The corresponding angles are given with
\begin{equation}\label{eq:mirror_angles}
 \phi^{'}_1=-\phi_1 ; \;
 \phi^{'}_2=-\phi_2 ; \;
 \phi^{'}_3=-\phi_3 .
\end{equation}
This ground state is related by the mirror symmetry
to the state defined with Eq.~(\ref{eq:angles}).

The high symmetry of the hexagon molecules makes it difficult
to perceive the ordering pattern presented in
Fig.~\ref{fig:patterns}(a). A more complicated task of finding
an orientational defect in this pattern becomes unfeasible.
Therefore for the purpose of visualization, we implement
a vector representation of hexagon
molecules shown in Fig.~\ref{fig:patterns}(b).
The vector rotation angle is three times a hexagon
rotation angle: $\phi^v_{lm}=3\phi^h_{lm}$. A vector can be rotated
from 0$^\circ$ to 360$^\circ$.
As a result, a difference between sublattices appears
to be more clear.

\begin{figure}[t]%
   \centering
   \includegraphics[width=\figurewidth]{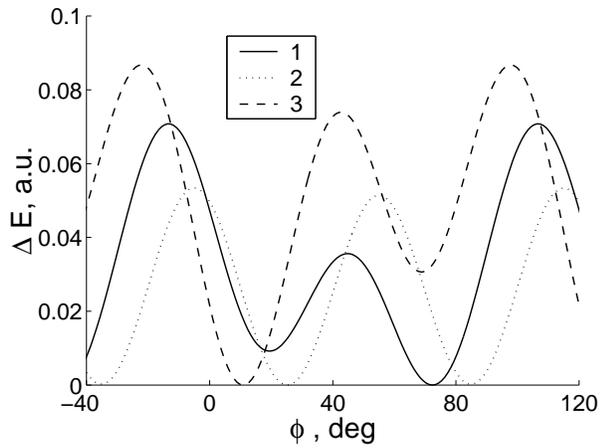}
   \caption{\label{fig:regular}
   Orientational potential profiles for regular molecules
   belonging to three different sublattices.}
   \end{figure}

Figure~\ref{fig:regular} shows a change of the molecule
interaction energy with a change of its orientation
for the three molecules belonging to three different sublattices.
It is clearly seen that the molecules are not identical.
All of them have double-well energy profiles, but the height of
the interwell energy barrier varies by a factor of 2.
The potential minima of the 2nd molecule are almost
energy degenerate, the energy difference being only 1/20 of the
barrier height (situation, similar to the case of fullerite).

\begin{figure}[t]%
  \centering
   \includegraphics[width=\figurewidth]{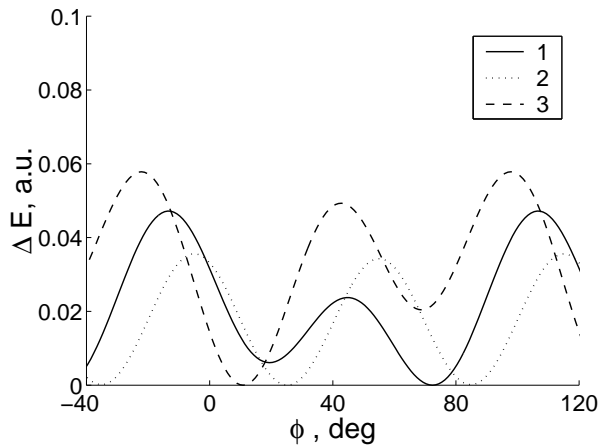}
   \caption{\label{fig:edge}
    Orientational potential profiles for  the edge molecules
    belonging to different sublattices.}
   \end{figure}

For a comparison, the same potential profiles are shown for the
molecules from the edge of the simulated lattice (see
Fig.~\ref{fig:edge}).
Such molecules keep only 4 of the 6 nearest neighbours
(molecules from two different sublattices are missing).
As a result, the overall potential profile is
lowered by a factor of 6/4.
The C$_{60}$ molecule at the fullerite (111) edge surface
is missing 3 neighbours from 3 different sublattices.
Therefore one could expect lowering the orientational barriers by
a factor of 12/9.\footnote{%
Nevertheless, it should be emphasized that a change (or
relatively weak lowering) of inter-molecular rotational barriers
appears to be too small for all the cases of the regular structure
to allow for an orientational tunnelling which is necessary for a
number of physical phenomena.
One has to remember that the mass of the C$_{60}$ fullerene
molecule is 720~a.u. It makes very strong constraint for the height
and the width of energetic barriers which are able to give an
observable probability (or frequencies \cite{TSB98})
of orientational tunnelling transitions.
}

But the real situation is even more complicated.
A three-dimensional character of fullerite lattice leads to
subdivision of the neighbours of an arbitrary bulk fullerene
molecule into only two categories,
denoted here as double-bond (to which the molecule is oriented
with the double bond), and pentagon (to which the molecule is
oriented with the pentagon or hexagon) neighbours.
The six double-bond neighbours belong to the (111) plane normal to the
molecule fixed $C_3$~axis. The rest six pentagon neighbours,
which give a major contribution to the molecule orientational
profile, are located in other (111) planes.

Therefore an edge molecule with a fixed $C_3$~axis normal to the
edge surface misses three pentagon neighbours, while the
molecules with three other directions of allowed rotation axis
are missing two double-bond and one pentagon neighbour each.
As a result, the potential relief of a molecule with a normal
rotation axis is more shallow than the relief of other molecules.
In this way, the molecules from the four different sublattices
which are identical by their rotational properties in the bulk
fullerite
become non-identical at the edge surface crystal defect due to a
loss of the symmetry.
This non-identity evidently reveals itself in the presence of two
additional lower temperature order-disorder phase transitions
reported in \cite{LPHLT01}.

\section{Linear orientational defects}

A general kind three sublattice two-dimensional triangular lattice
allows for orientational ordering of three different types.
Molecular orientations for these ordering types are related
to each other with cyclic permutations of the rotation angles
$\phi_i$ ($i=1,2,3$, cf. Eq.~(\ref{eq:angles})) for the molecules
located at the vertices of a lattice triangle
(eg., a triangle 123 shown with a solid line in Fig.~\ref{fig:patterns}).
In the case of the considered hexagon molecules an existence
of the mirror orientational twin defined with
Eq.~\ref{eq:mirror_angles} leads to appearance of three additional
ordering types, which are related to the basic permutation
ones with the mirror symmetry.

As a result, the considered lattice
allows for simultaneous existence of orientational domains with
6 different ordering types.
A boundary between two domains contains a linear orientational
defect (domain wall). Such defect can involve a permutation
(clockwise or counterclockwise) or a
mirror transformation (with a center at 3 different lattice sites)
of molecular orientations.\footnote{%
For the case of fullerite, there are 4+4 different ordering
types and 3+4 different inter-domain boundaries (not related with
the symmetry operations).}

\begin{figure}[t]%**
  \centering
   \includegraphics[width=\figurewidth]{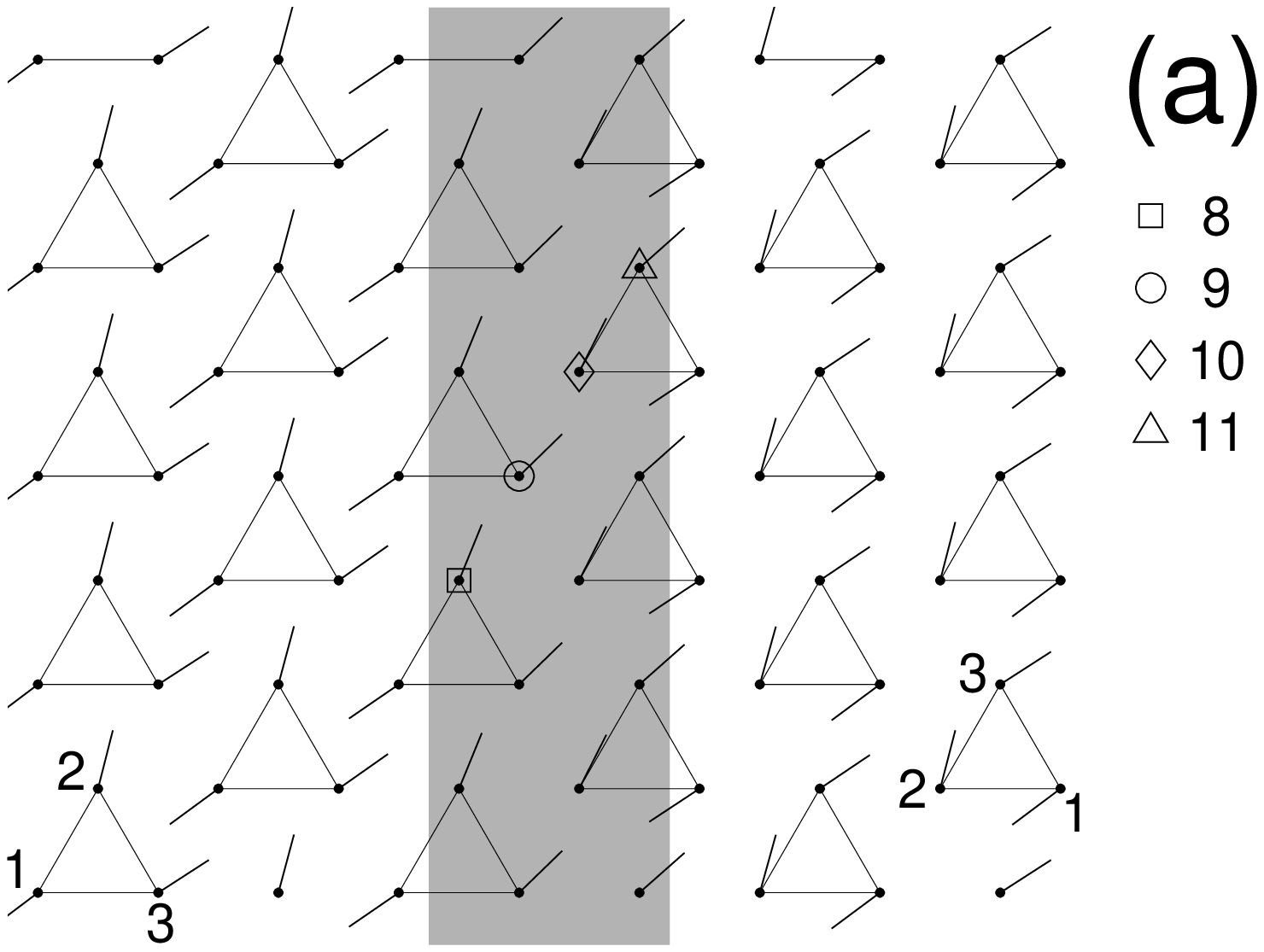}
   %\hfill
   \includegraphics[width=\figurewidth]{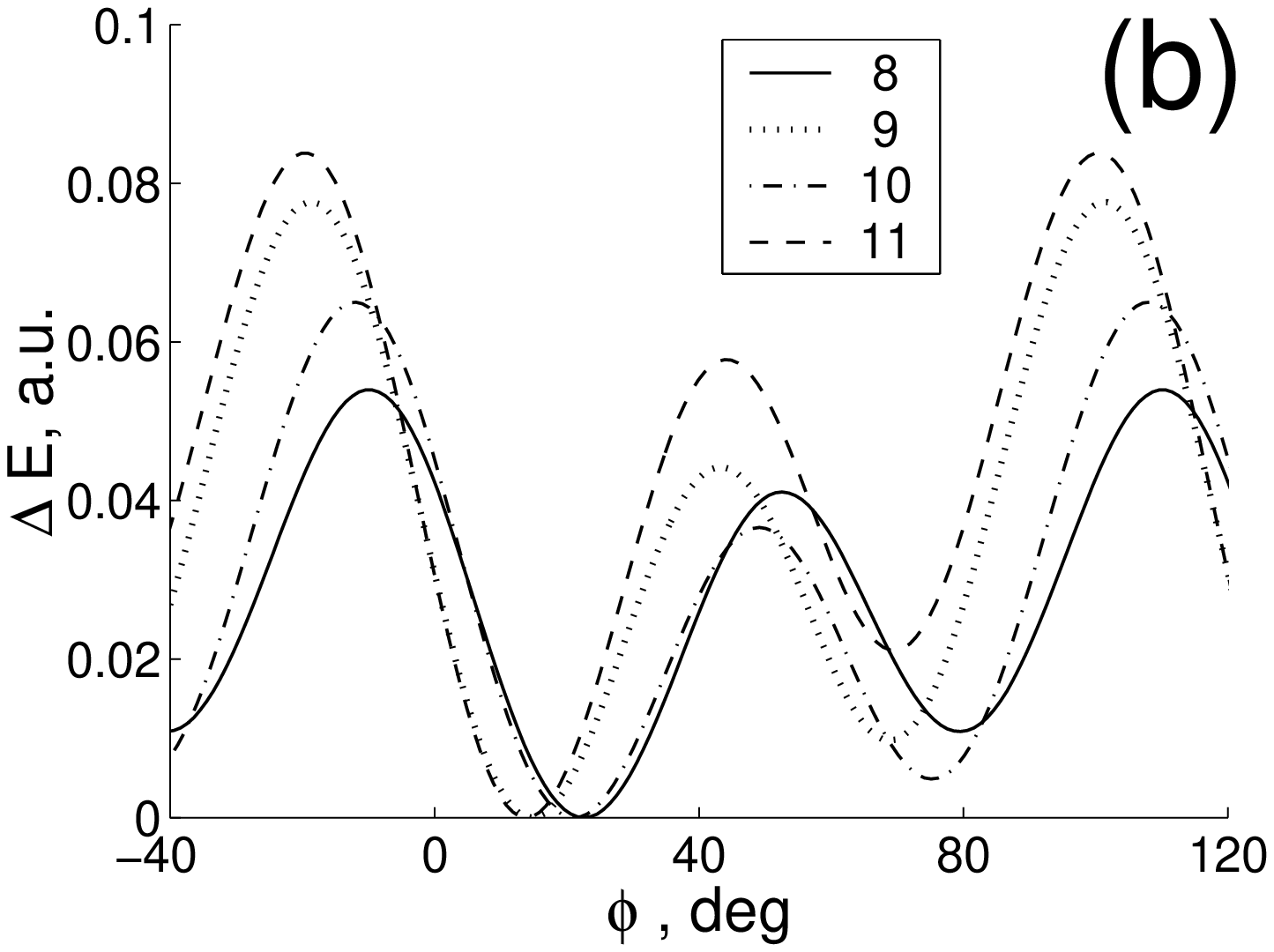}
   \caption{\label{fig:perp}
   Permutation domain wall~(a) perpendicular to close-packed row,
   and orientational potential profiles~(b) for the
   four marked molecules identified with the lattice index $m$
   ($l$=10).}
   \end{figure}

A domain wall of the permutation type is presented in
Fig.~\ref{fig:perp}(a).
The rotation angles of the molecules located at the vertices
of a lattice triangle
(shown with solid lines)
have the values $\phi_1$, $\phi_2$, and $\phi_3$
in the left  domain.
In the right domain they are equal to $\phi_2$, $\phi_3$, and
$\phi_1$, correspondingly.
The domain wall (grey) is relatively narrow.
Its width (measured along the horizontal close packed $l$ direction)
is about one period of 3-sublattice structure.
As seen along the close packed $m$ direction,
this defect can be regarded as obtained by removal of one element
from an ideal sequence $\ldots1231231\ldots$ of molecular
orientations. The resulting sequence is $\ldots123|231\ldots$ .

Orientational dependence of the potential energy of the four
central molecules from the domain wall is given in
Fig.~\ref{fig:perp}(b).
The molecules are marked in Fig.~\ref{fig:perp}(a) and
labeled with their $m$ index, while $l$ is taken to be 10.
Molecules 8 and 10 have the orientations of the type 2,
and the rotation angles of molecules 9 and 11 are close to
$\phi_3$.
The potential profiles are quite close in the form to the profiles
of the regular molecules (shown in Fig.~\ref{fig:regular}),
but one of the two potential barriers is lowered for each
molecule.

\begin{figure}[t]
  \centering
   \includegraphics[width=\figurewidth]{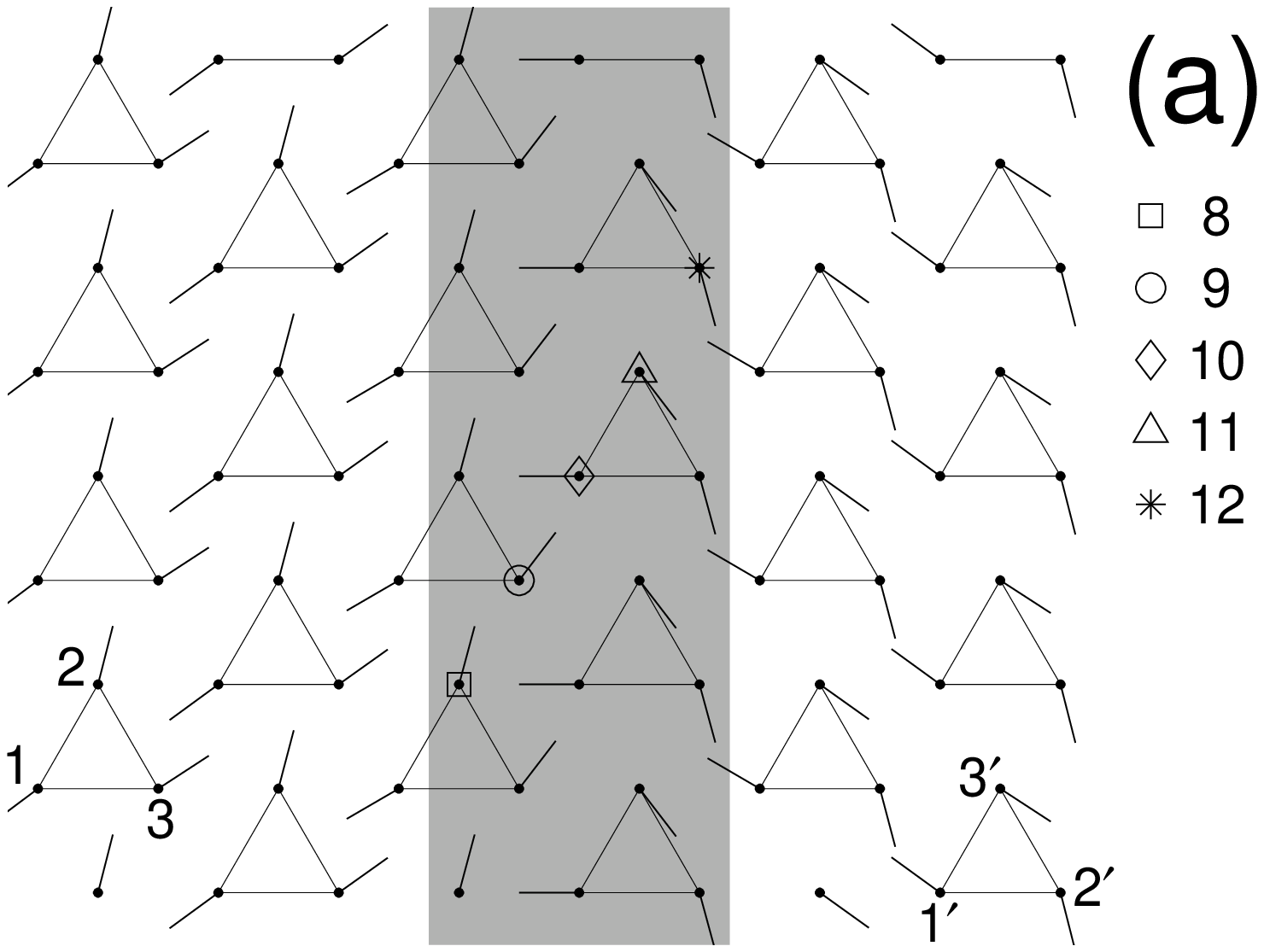}
   \includegraphics[width=\figurewidth]{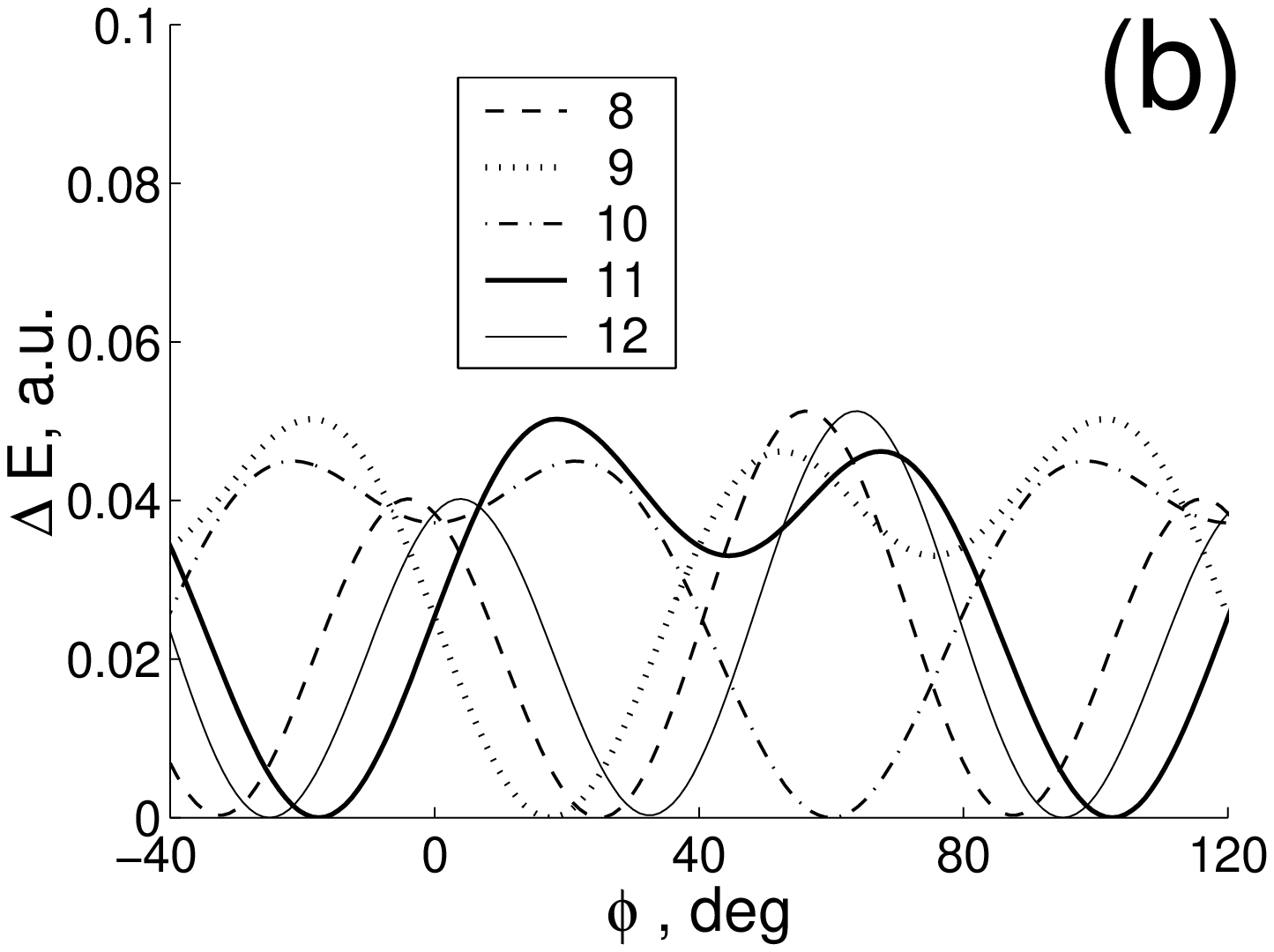}
   \caption{\label{fig:perp-mirr}
   Mirror domain wall~(a) perpendicular to close-packed row,
   and the potential profiles~(b) for the five marked molecules
   with $l$=10 and with the indicated $m$ value.
   }
   \end{figure}

Orientational domain walls of a mirror nature
are wider than the permutation ones.
Fig.~\ref{fig:perp-mirr}(a) gives an example of the mirror
domain wall. For this wall,
a sequence of molecular orientations in the $m$ direction is
$\ldots123?3^{'}2^{'}1^{'}\ldots$, where a question mark stands
for a molecule in the mirror plane.
This molecule does not fit any regular orientation.
Instead, it reflects the mirror symmetry of the wall.
Figure~\ref{fig:perp-mirr}(b) clearly shows the
orientational potential minimum
of the molecule 10 to be located
at the rotation angle $\phi=60^\circ$.
Such an orientation corresponds to aligning one of the mirror
planes of the hexagon molecule to the domain wall mirror plane.

The mirror symmetry of orientational defect is also manifested
through the symmetry of potential curves of other molecules.
The potential profiles of the molecules 9 and 11 (orientations 3
and 3$^{'}$), 8 and 12 (orientations 2 and 2$^{'}$) are related
through $\Delta E_9(\phi)=\Delta E_{11}(-\phi)$ and
$\Delta E_8(\phi)=\Delta E_{12}(-\phi)$, respectively.

In the vector pattern of Fig.~\ref{fig:perp-mirr}(a),
this symmetry is given with the clockwise-counterclockwise
vector rotations on the two different wall sides.
Since the rotation angles are measured form the X direction,
the vectors representing molecule 9 and 11 (8 and 12) rotations
are related with the mirror plane parallel to the X direction.

The effect of the domain wall on the potential relief of the
molecules 8 and 12 (orientations 2 and 2$^{'}$) is found
to consist in a
slight lowering of one of the two barriers. For the molecules
9 and 11 (orientations 3 and 3$^{'}$) close to the center of the
domain wall, both the potential barriers are lowered considerably.
But the position and the relative height of the secondary minimum
are unchanged, resulting in a shallow character of this minimum
seen in Fig.~\ref{fig:perp-mirr}(b).

\begin{figure}[t]
\centering
\includegraphics[width=\figurewidth]{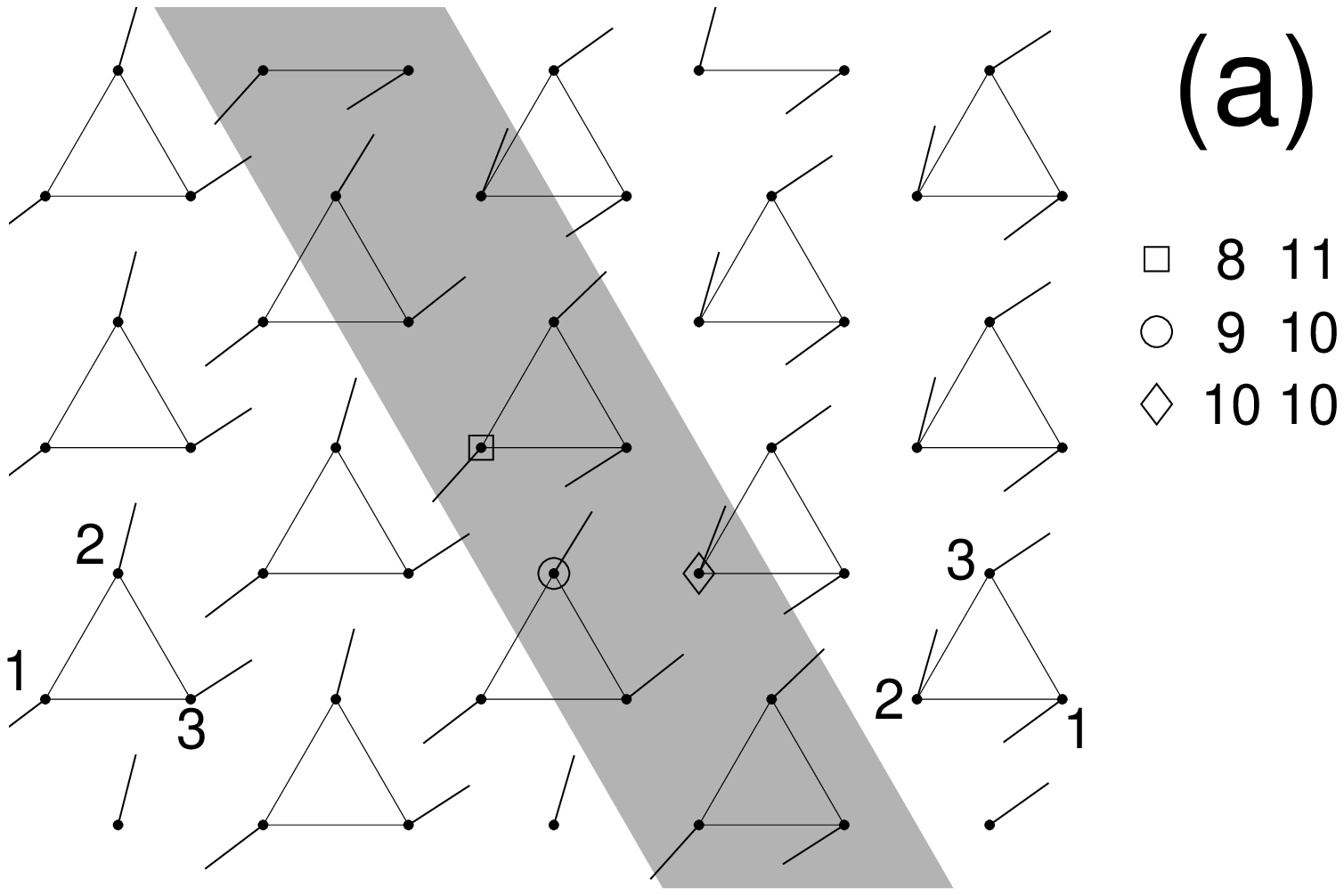}
\includegraphics[width=\figurewidth]{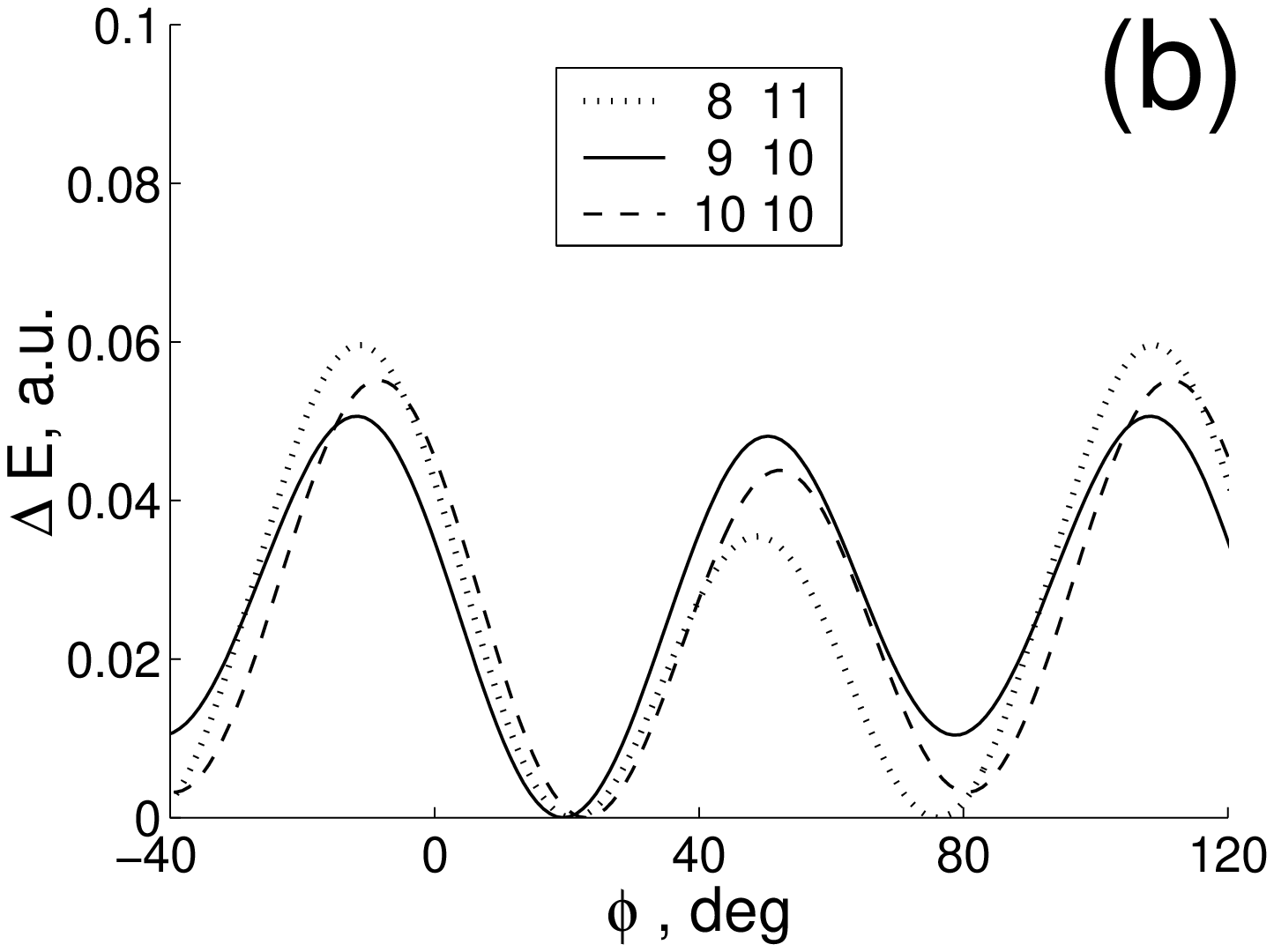}
\caption{\label{fig:par}
Permutation domain wall parallel to close-packed row~(a),
and orientational potential profiles~(b) for the marked
molecules with the given lattice indexes $l$ and $m$. }
\end{figure}

The domain walls given in
Figs.~\ref{fig:perp}, \ref{fig:perp-mirr}
have their directions parallel to one of the sublattice period vectors,
and perpendicular to one of the close-packed molecular
row directions. At the same time, there is a possibility for a domain
wall to lie along the close-packed molecular rows.
An example of a permutation domain wall of this kind
is presented in Fig.~\ref{fig:par}(a).
The relationship between the left and right domains
here is the same as in Fig.~\ref{fig:perp},
but the location of the domain wall line is different.
As a result, the molecular orientation sequence
along the $m$ molecular row can now read
$\ldots231|312\ldots$ ($l=8$),
$\ldots312|123\ldots$ ($l=9$), or
$\ldots123|231\ldots$ ($l=10$).
Therefore, the central part of the domain wall contains
molecules with 6 different
potential profiles (orientations 1, 2, and 3 from the left domain,
and orientations 3, 1, 2 from the right domain).
The three potential profiles with the lowest energy barriers
are shown in Fig.~\ref{fig:par}(b). Noteworthy that here we gain
a low barrier profile with almost energy degenerate minima
(see dotted curve).

\begin{figure}[t]
\centering
\includegraphics[width=\figurewidth]{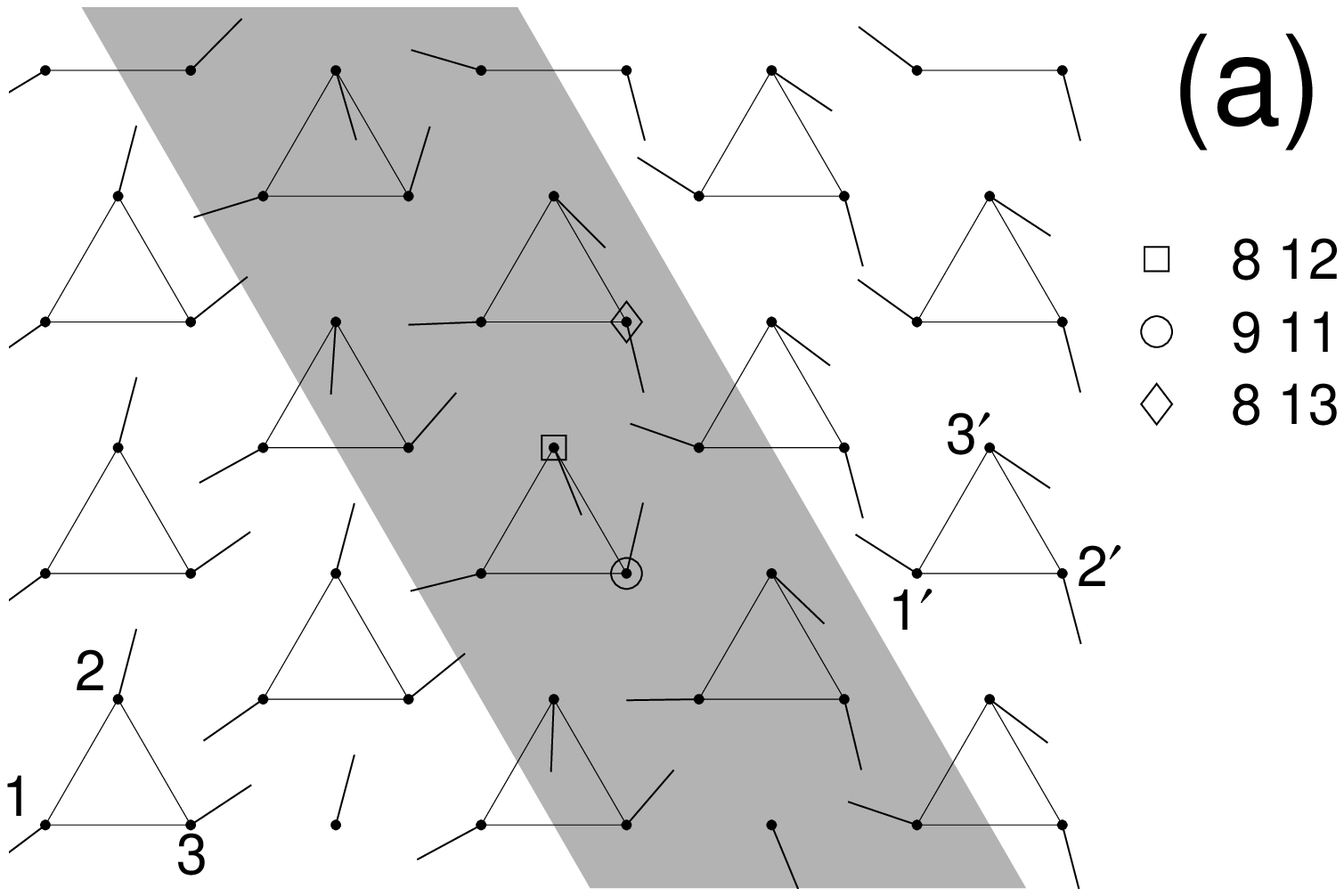}
\includegraphics[width=\figurewidth]{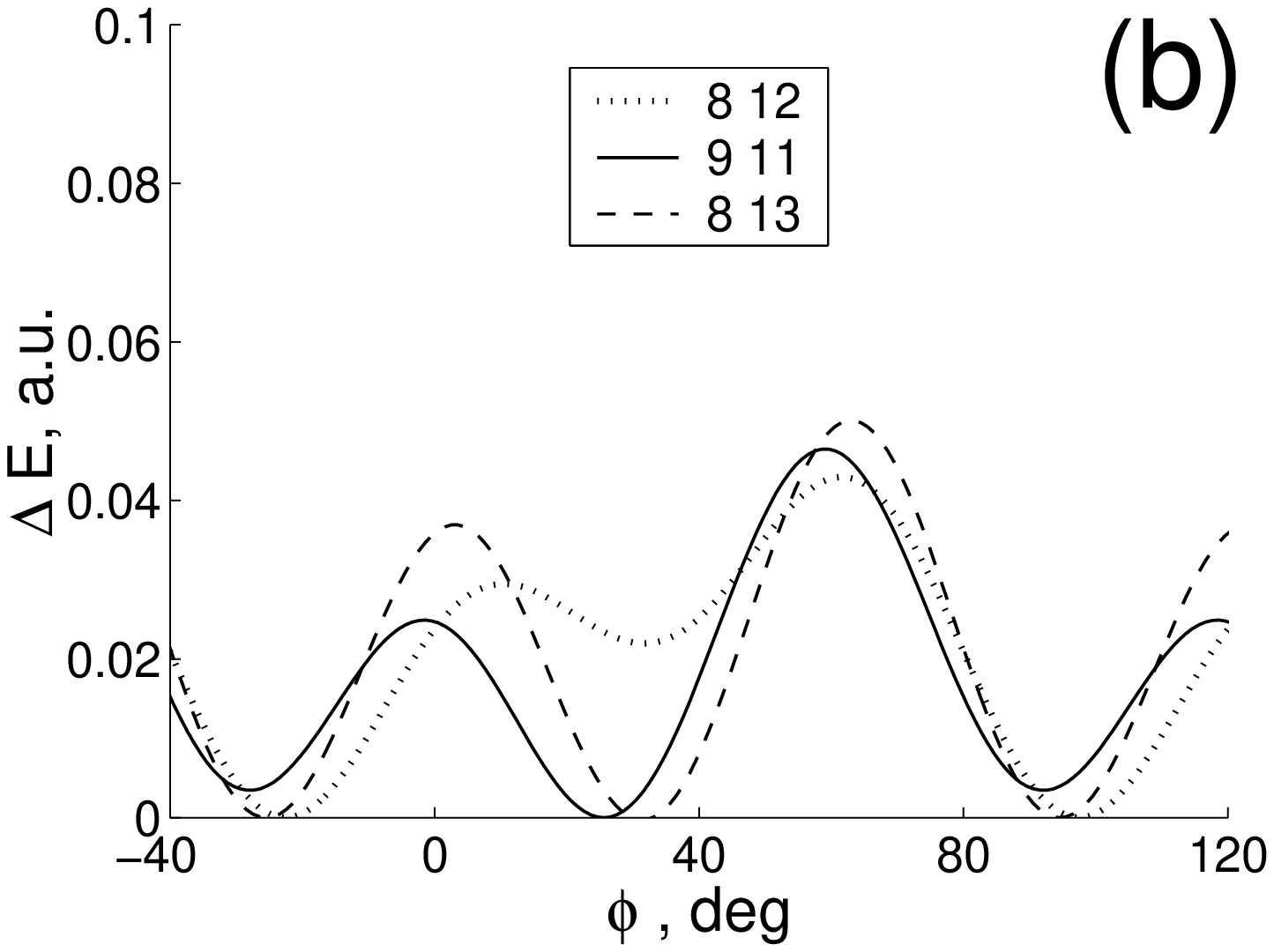}
\caption{\label{fig:par-mirr}
Mirror domain wall parallel to close-packed row~(a),
and the potential profiles~(b) for the marked
molecules (lattice indexes $l$ and $m$ are indicated).}
\end{figure}
A mirror nature domain wall parallel
to close-packed molecular row
has a more complicated structure shown in
Fig.~\ref{fig:par-mirr}(a).
It is again wider than the permutation wall,
so that the molecules from {\em three}
close-packed rows have substantially corrupted
orientational potential relief.
As a result, the number of
intra-wall molecules with different orientational profiles
increases up to 9, opposed to 6 different profiles
for a permutation wall.
Furthermore, the direction of the domain wall
does not coincide with the lattice mirror plane,
so there are no mirror symmetry in the pattern of
Fig.~\ref{fig:par-mirr}(a), and, respectively,
no symmerty relations for the potential curves
(conf. the mirror symmetry of the potential profiles
shown in Fig.~\ref{fig:perp-mirr}(b) for the domain wall
perpendicular to close-packed molecular row).
In Fig.~\ref{fig:par-mirr}(b) we give the orientational
potential profiles for the three molecules with the
lowest interwell energy barriers.
It is seen that there exists a molecule (o)
which interwell
barrier is about 1.4 times lower than the lowest
of the regular molecules interwell barriers.
The molecule is situated in the center of the domain wall
and marked with a circle.
The corresponding potential profile is plotted with
a solid line in Fig.~\ref{fig:par-mirr}(b).
The obtained reduction of the orientational
interwell barrier is caused
by a less correlation between
the nearest neighbour molecules
(every molecule within the considered wall has neighbours
of six different orientations).

\section{Two-dimensional defects}
The results on the modelling of the straight domain walls
in the considered system show that the molecules with the
most shallow potential profile tend to appear at sites
with the reduced correlation between the orientations of
the neighbour molecules. For the straight walls such
condition is met at the boundary of  two domains
with  different sets of equilibrium molecule orientations
(mirror domain walls).

Then it is straightforward to continue
the search for the shallow
potential molecules in the core regions of essentially
two-dimensional orientational defects.
One of such promising two-dimensional defects
is a meeting point of three different domains.
Molecules at the center of this defect should
have three pairs of neighbours belonging to three
different domains,
so one could expect for an additional decrease
of interwell barriers with respect to the two-domain
boundary case.

The results of numerical calculations indeed show
the further reduction of interwell potential barriers
at the boundary of three orientational domains.
The most effective reduction is found to take place
in the presence of mirror boundaries.

  \begin{figure}[t]%**
  \centering
   \includegraphics[width=\figurewidth]{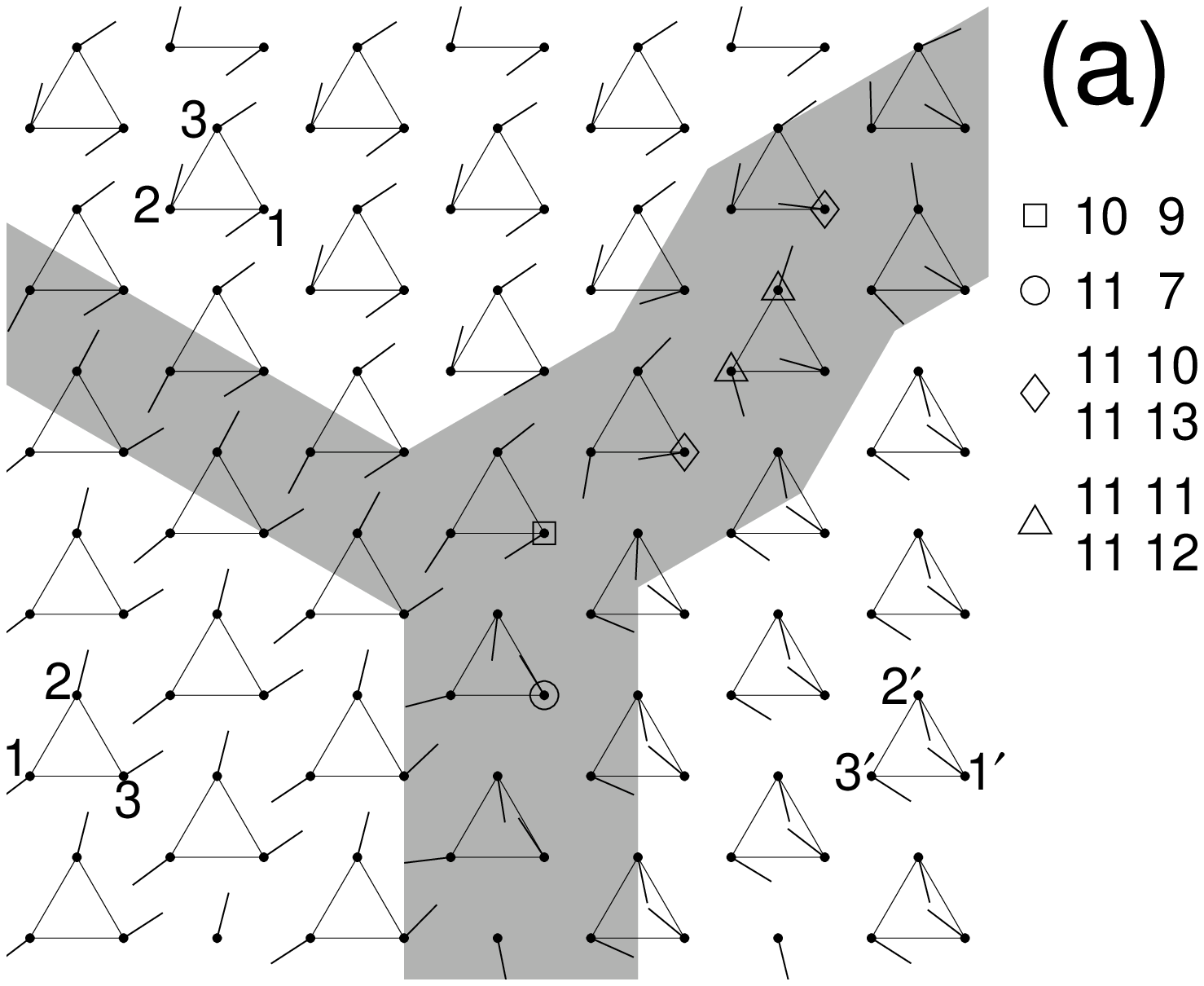}
   \includegraphics[width=\figurewidth]{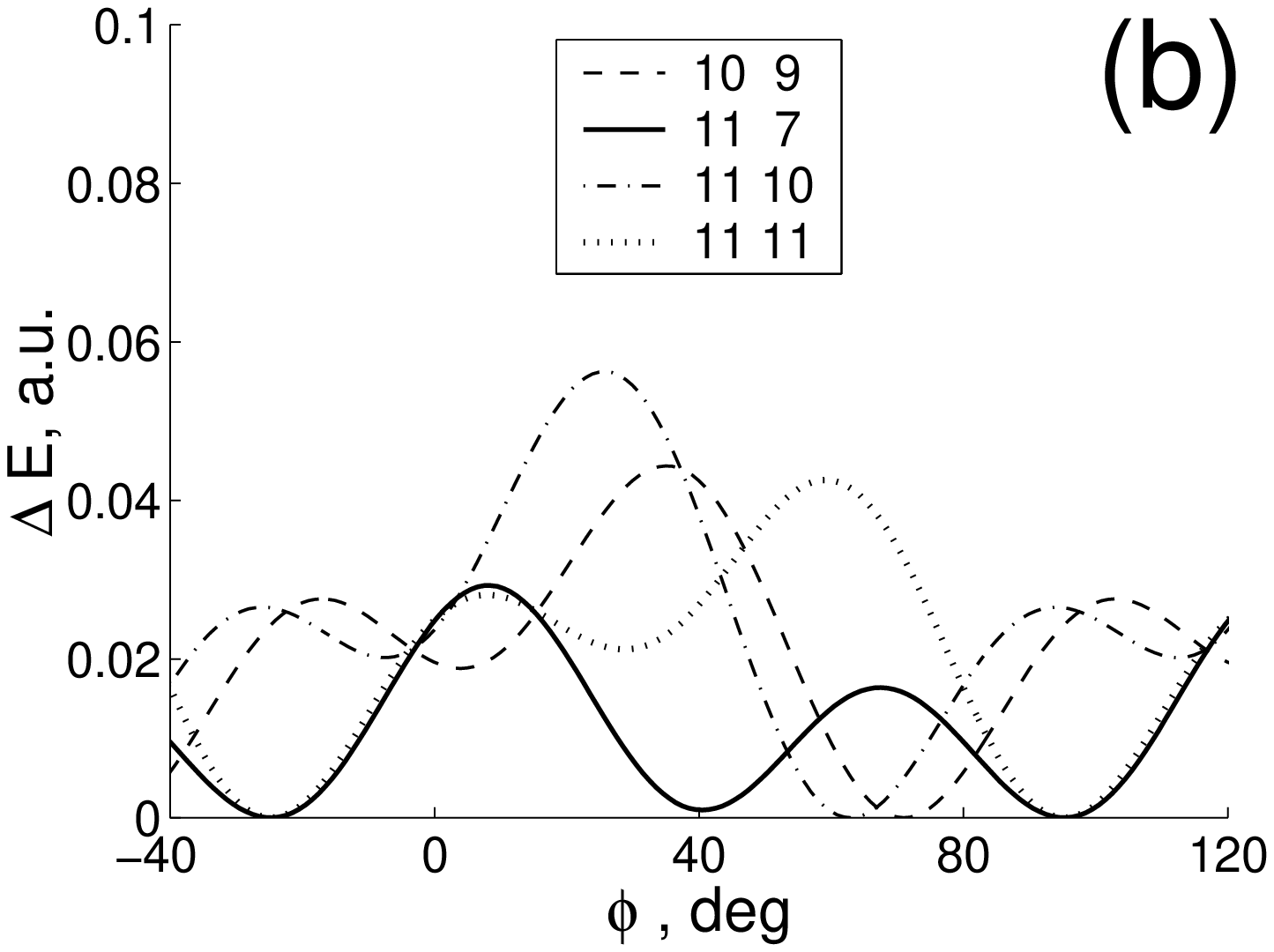}
   \caption{\label{fig:three}
   Structure of orientational defect~(a)
   formed at the boundary of three orientational
   domains, and the potential profiles~(b)
   for some chosen molecules in the defect core region.
   Pairs of molecules marked with the same sign
   ($\triangle$~or~$\Diamond$)
   have the symmetry related potential profiles,
   so only one of the profiles is given for each pair.
   Molecules are labeled with $l$ and $m$ indexes.}
   \end{figure}
Figure~\ref{fig:three}(a)
shows an example of orientational defect formed
at the intersection of three domain walls perpendicular to
molecular rows.
The left (narrow) domain wall is of a permutation
type, the other two (the bottom one and the right one) have
a mirror nature and are much wider.
The right domain wall incorporates a kink in order
to minimize a surface spanned by the defect.
Molecules with the lowest interwell barriers are marked.

As it is mentioned above, no significant potential barrier
reduction has been observed for straight domain walls
perpendicular to close-packed molecular rows. Therefore
the marked molecules can be seen only at the crossing
of the three walls.
The corresponding orientational potential profiles
are plotted in Fig.~\ref{fig:three}(b).

It is surprising that the potential profile
with the least energy barrier belongs not to
the molecule~$\Box $ situated in
the very center of the defect (potential curve plotted
with a dashed line) with totally different orientations
of all the 6 nearest neighbours, but to the molecule
located at the beginning of the bottom domain wall
(o, solid line).
For the last molecule the orientations of the nearest neighbours
differ only slightly from that in the straight wall,
but the interwell potential barrier is 2.3 times lower
than the lowest regular molecule barrier.

The other four molecules which are marked in
Fig.~\ref{fig:three}(a) are located within the center
of the kink in the right domain wall.
At a closer look, one can find a kind of a symmetry center
at the middle of the line between the molecules marked with
$\triangle$. An exact symmetry is following:
if the centers of two molecules are related with
inversion symmetry, these molecules have the rotation angles
which have equal absolute values, but different signs.
Therefore the two molecules marked with $\triangle$
(as well as the two molecules marked with $\Diamond$)
have the same orientational dependence of
intermolecular interaction potential, the only difference
being in the clockwise or counterclockwise direction
of molecule rotation.
This can be compared to the symmetry of potential curves
in Fig.~\ref{fig:perp-mirr}(b), but there is no mirror
plane in the present case.
To avoid having a very complicated picture,
only one of the two symmetry related curves
is shown in Fig.~\ref{fig:three}(b) for each pair
of molecules. Both the dotted and the dash-dotted
curves have an interwell energy barrier which is less
than the lowest energy barrier characteristic for
regular molecules.
This means that at the center of the kink in a domain wall
(also a two-dimensional defect) molecules have
ill-correlated nearest neighbours.
Therefore the case of a kinked domain wall
has to be investigated more thoroughly.

   \begin{figure}[t]%**
  \centering
   \includegraphics[width=\figurewidth]{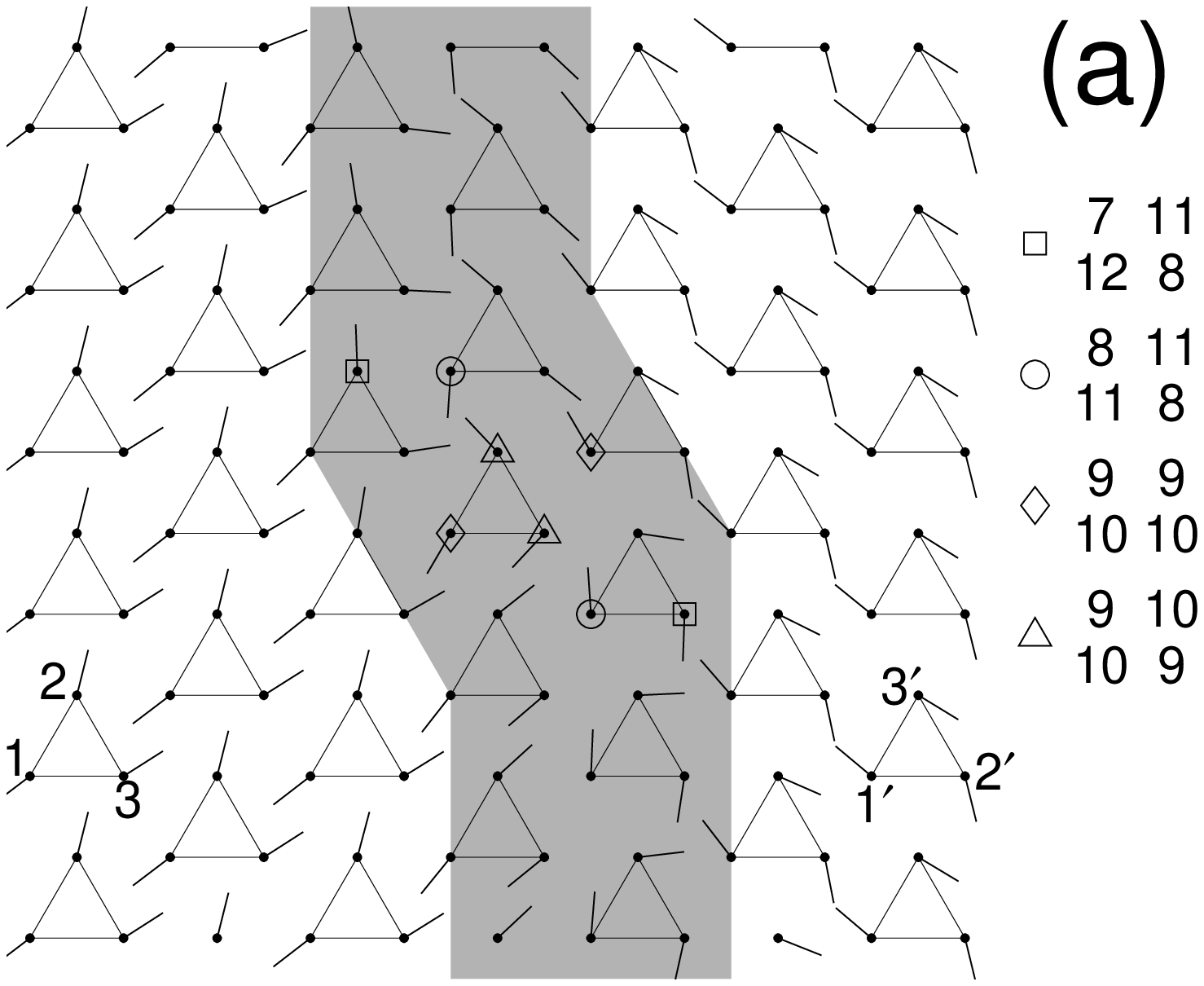}
   %\hfill
   \includegraphics[width=\figurewidth]{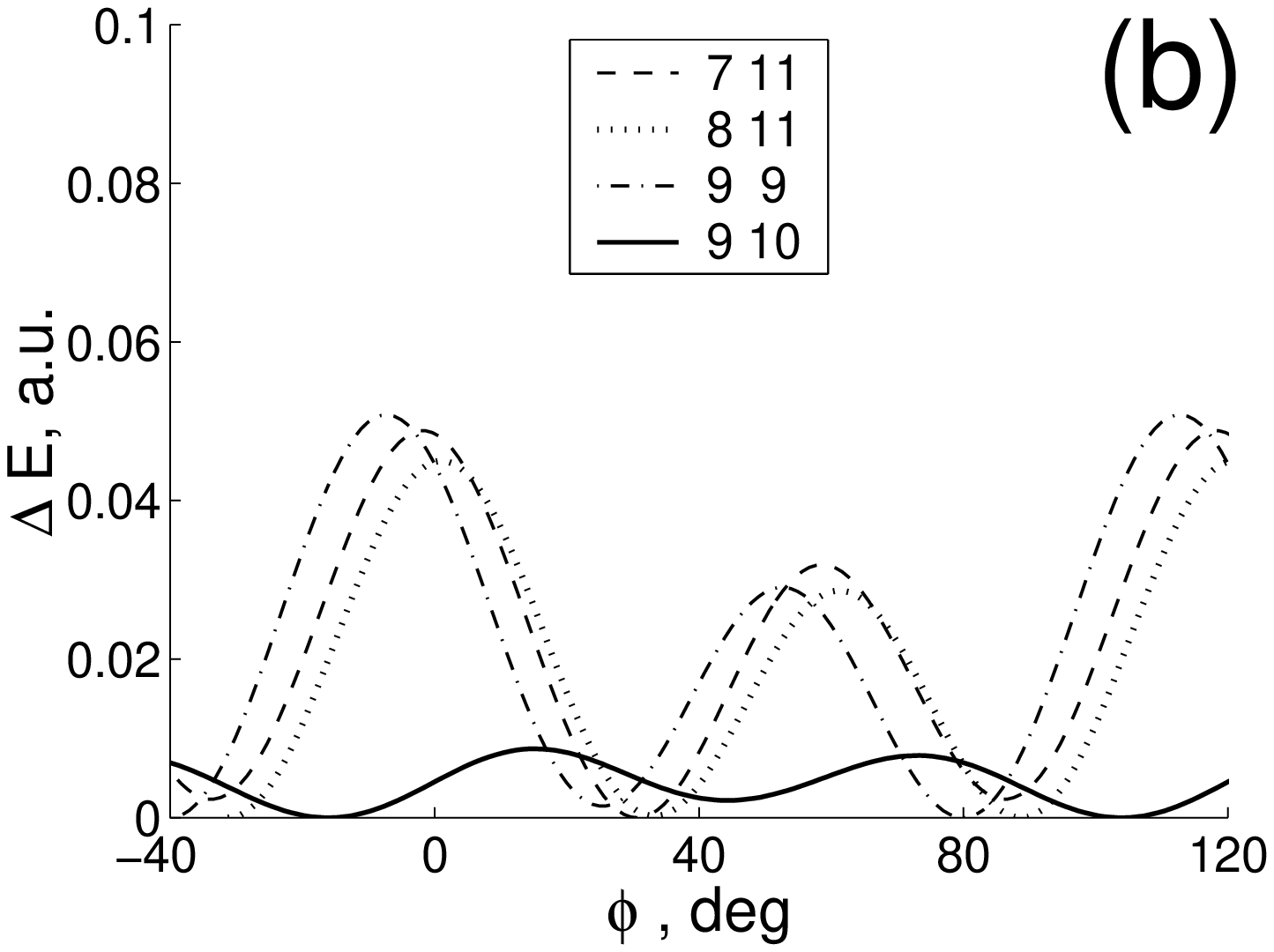}
   \caption{\label{fig:kink}
   The structure of the kink in the mirror domain wall
   (a), and orientational potential profiles (b) for
   marked molecules. Only one potential curve is given
   for every pair of symmetry related molecules
   which are marked with identical signs.
   Indexes l and m are indicated.
   }
   \end{figure}
Figure~\ref{fig:kink}(a) shows a structure of the kink
which contains the molecule with the lowest
height of the orientational interwell barrier
obtained in our simulations.
This molecule (in fact, two molecules,
since the kink has a center of symmetry
of the described above kind)
is located at the very center of the kink,
and the corresponding potential curve is shown
in Fig.~\ref{fig:kink}(b) with a solid line.
The height of the interwell potential barrier
is already 5 times less than for the case
of regular molecules.

\section{Totally uncorrelated neighbourhood
configuration}

The three-dimensional defect structure of
the real fullerite can be even more complicated.
As a result, some molecule can have the neighbours
which orientations are fixed by different
elements of the defect network.
In the frame of our simple two-dimensional model
such neighbourhood would be totally uncorrelated,
and the height of the interwell barriers could
be further lowered.
Therefore it is interesting to know a minimum
possible height of the molecule interwell potential
barrier for an arbitrary  orientational
configuration of its neighbour molecules.

\begin{figure}[t]
  \centering
   \includegraphics[width=\figurewidth]{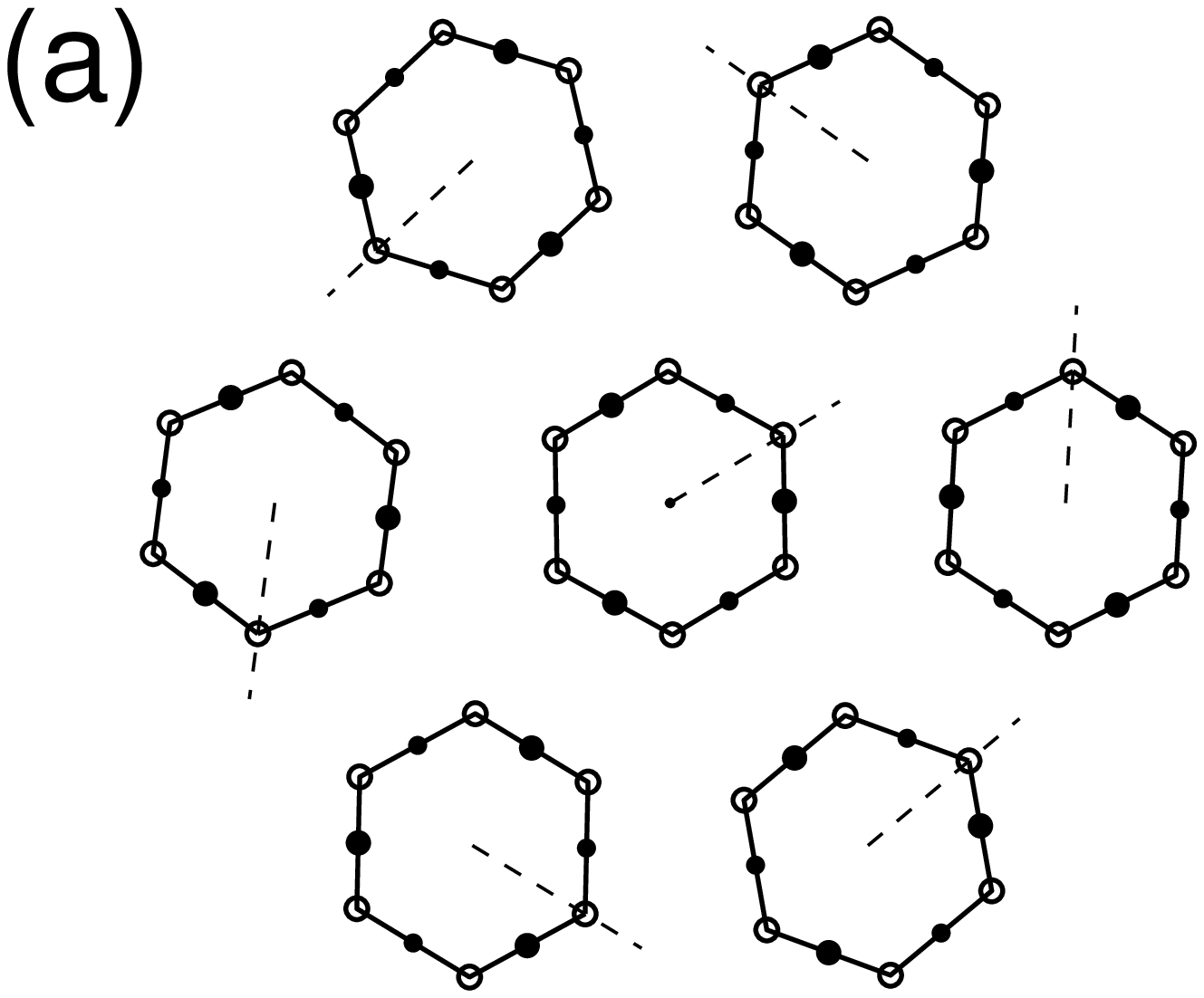}
   \includegraphics[width=\figurewidth]{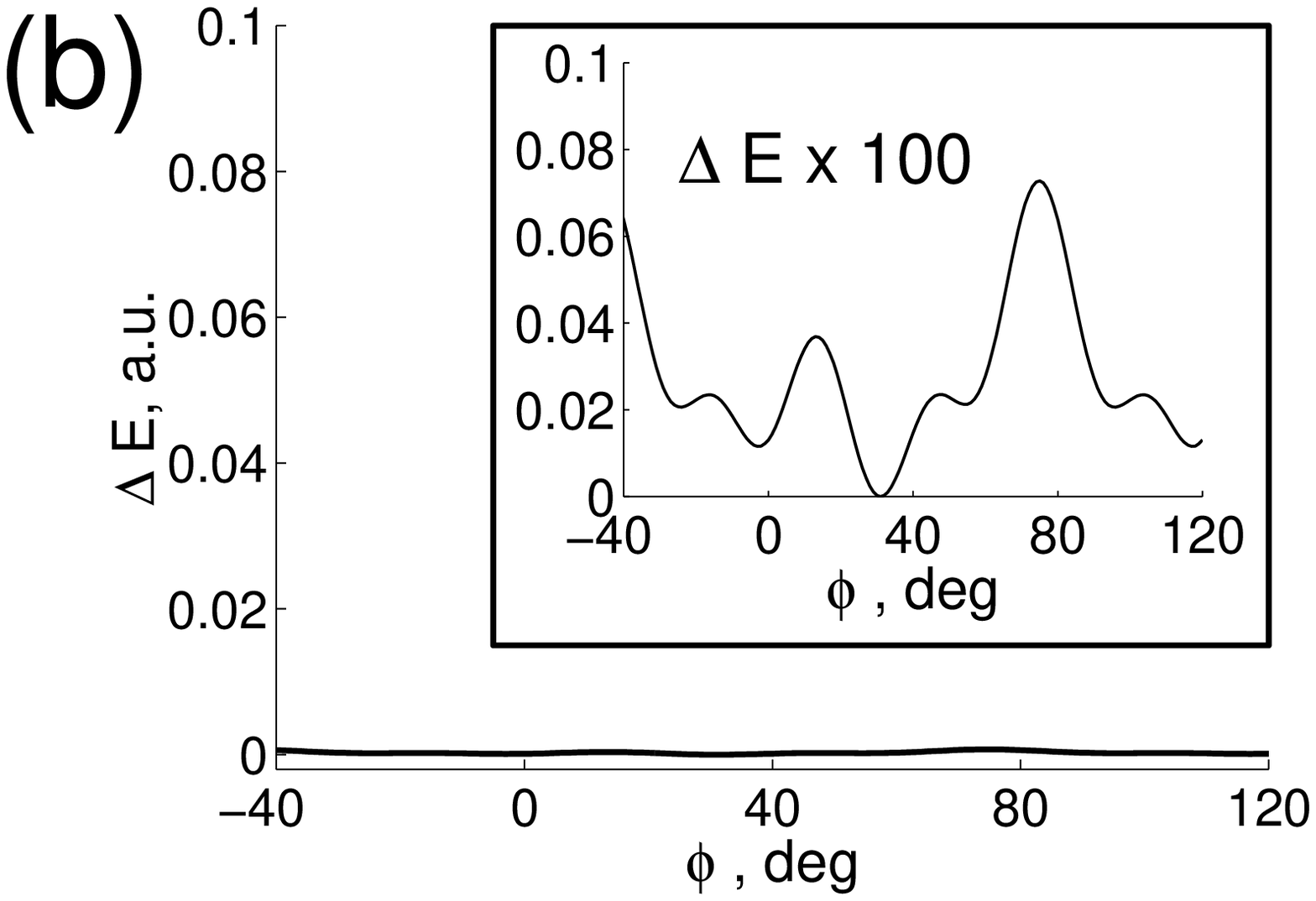}
   \caption{\label{fig:rand}
   A molecular configuration with nearly symmetric
   orientations of the outer molecules (a)
   and the corresponding shallow potential profile
   of the central molecule (b).
   An inset in the bottom panel shows a magnified
   potential curve.
   }
   \end{figure}

\begin{figure}[t]
   \includegraphics[width=\figurewidth]{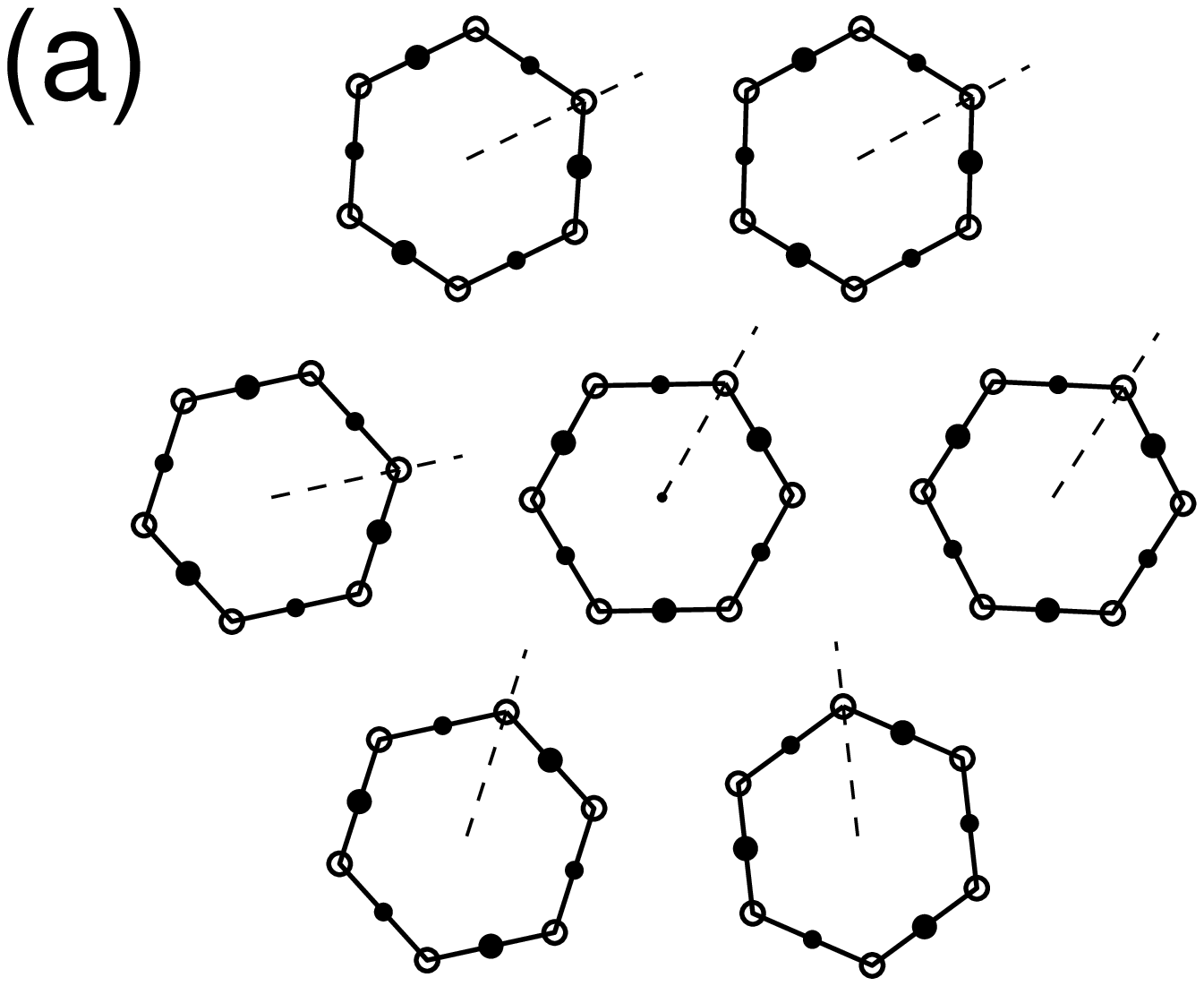}
   \includegraphics[width=\figurewidth]{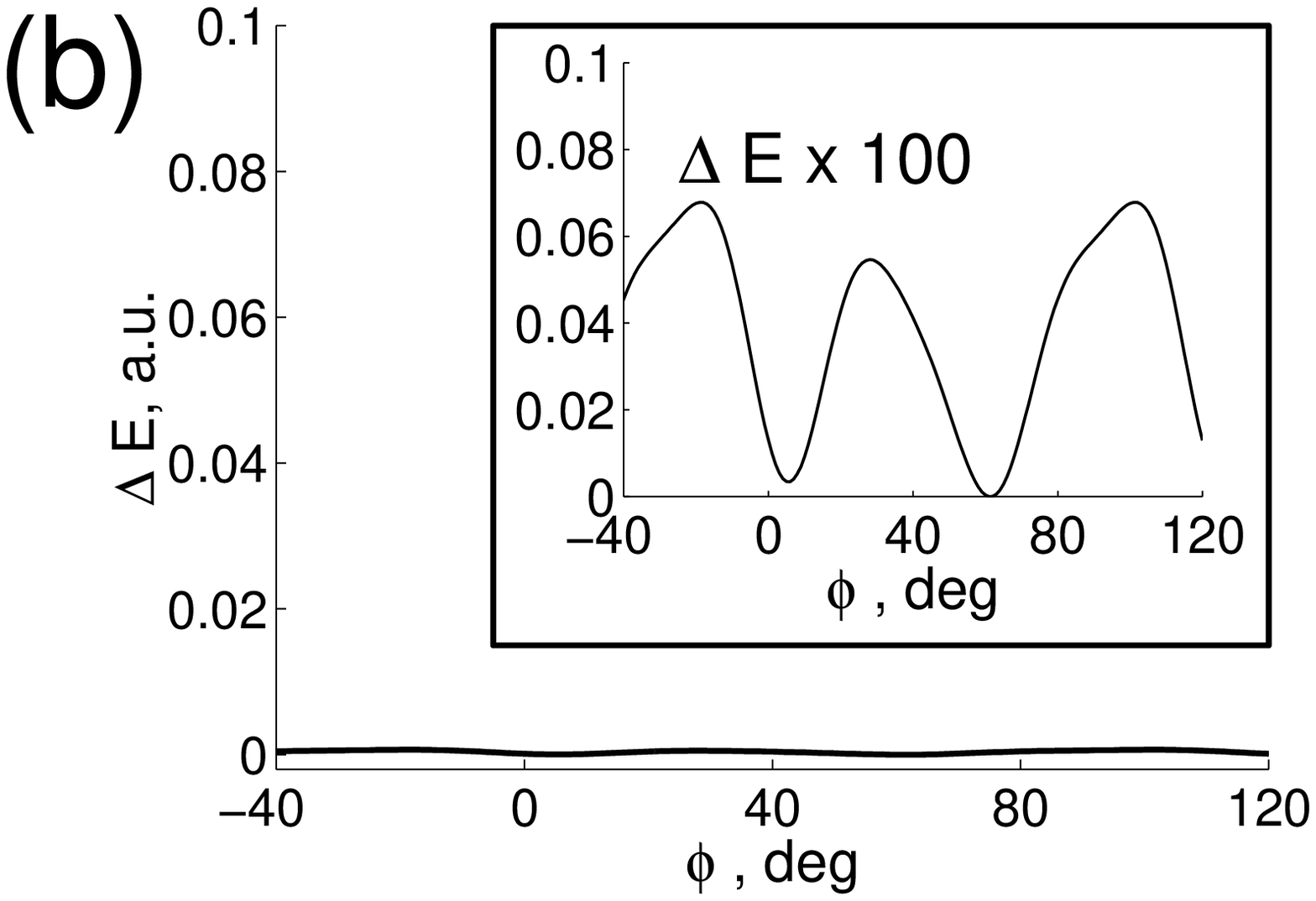}
   \caption{\label{fig:rand2}
   The same as in Fig.~\ref{fig:rand}:
   Another nearly symmetric configuration with a
   two-well potential profile.
   }
   \end{figure}

\begin{figure}[t]
   \includegraphics[width=\figurewidth]{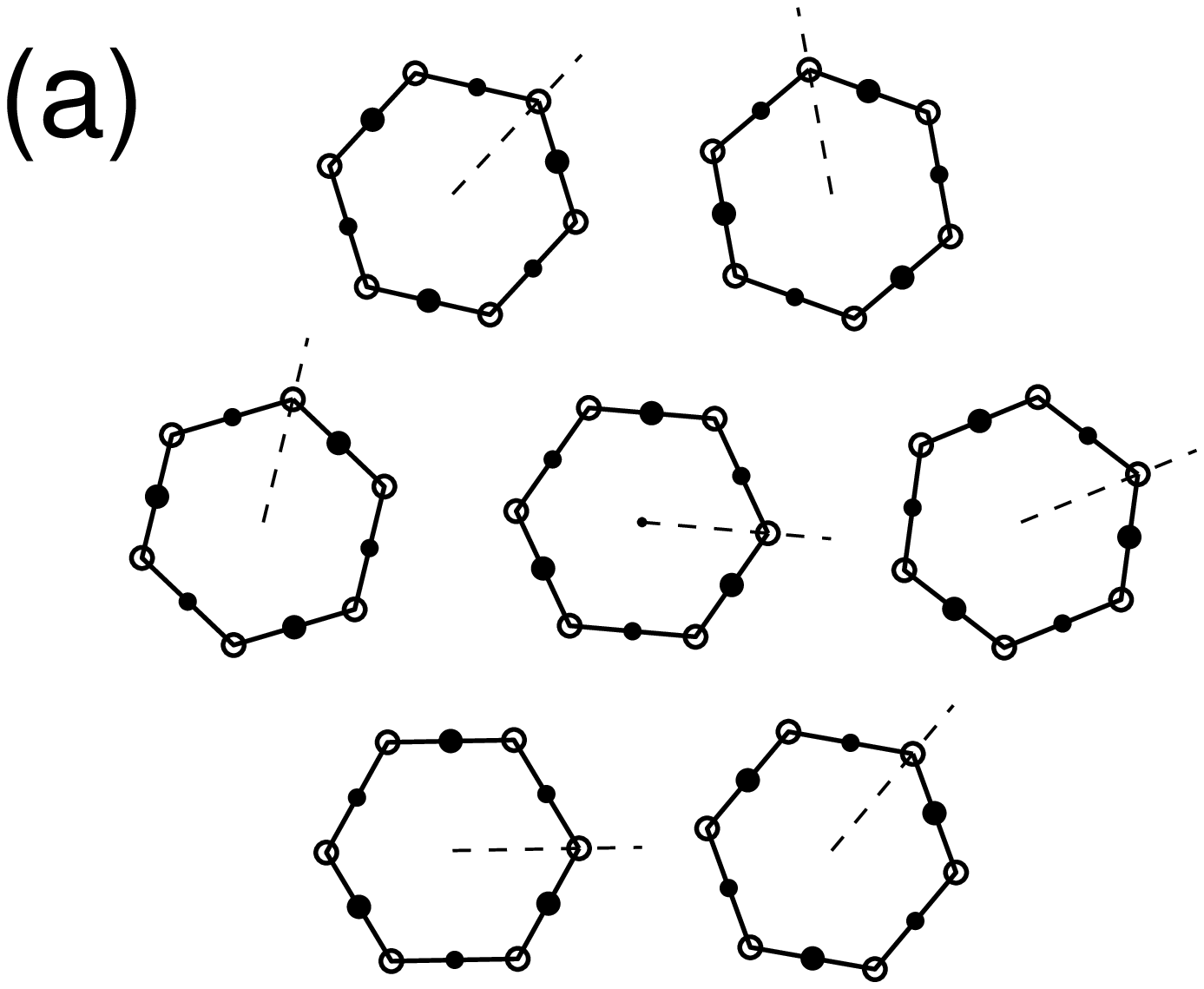}
   \includegraphics[width=\figurewidth]{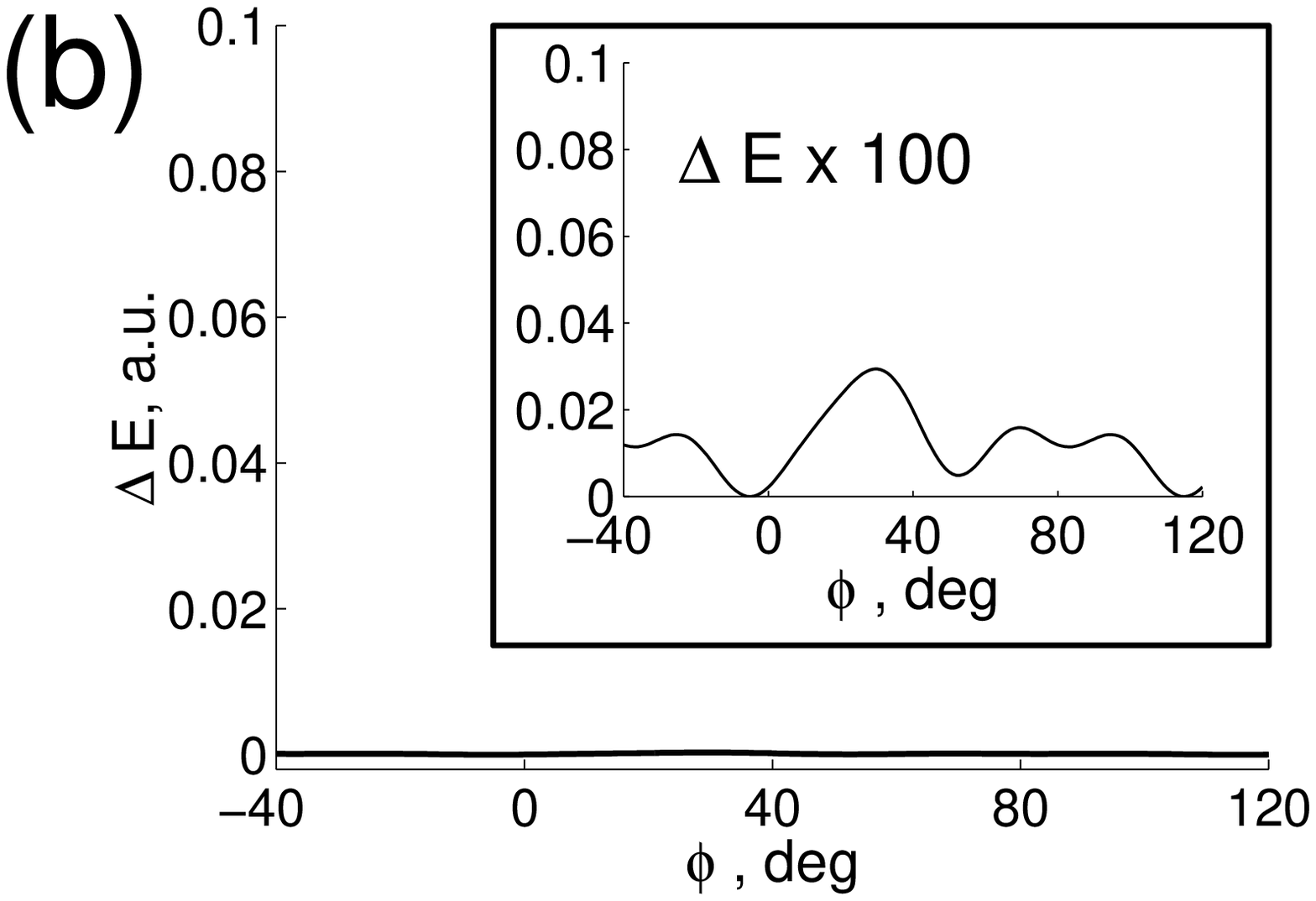}
   \caption{\label{fig:rand3}
   The same as in Fig.~\ref{fig:rand}:
   Nonsymmetryc configuration with the lowest
   found interwell barriers.
   }
   \end{figure}

For this purpose, let us consider a system of 7 hexagon
molecules located at the sites of hexagonal lattice,
so that one central molecule has 6 nearest neighbours.
Rotation angles of the outer molecules are fixed to
be equal to 6 random numbers between $0^{\circ}$ and
$120^{\circ}$, and then the orientational potential
profile of the central molecule is calculated.
Configurations with the most shallow potential profiles
obtained in the course of about $10^6$ different realizations
of random neighbourhood configuration are shown in
Figs.~\ref{fig:rand}, \ref{fig:rand2}, and \ref{fig:rand3}.

Figure~\ref{fig:rand} gives an example of a molecular
configuration with interwell potential barriers
of a central molecule reduced by
two orders of magnitude with respect to the case of
totally orientationally ordered lattice.
This configuration is nearly symmetric (the outer
molecules have rotation angles about $\pm 30^{\circ}$).
The central molecule has a four-well orientational potential
profile with the main minimum located close to $30^{\circ}$.
One could expect that a completely symmetric configuration
might have even more shallow potential profile of
the central molecule, because of the increase of the interaction
energy at the minima of the potential.
Contrary to the expectations, the exactly symmetric
configuration (not shown) has
an order of magnitude higher interwell barriers than the one
shown in Fig.~\ref{fig:rand}.
Thus, interwell barriers prove to be extremely sensitive
to even very smal rotations of the molecules.

The case of the molecular configuration with a two-well
orientational profile of the central molecule is presented in
Fig.~\ref{fig:rand2}. If one does not take into account
the difference between the values of negative charges,
this configuration seems to be close
to having a mirror symmetry.
Probably, namely this difference leads to an increase of
the interaction energy at the potential minima.

The molecular configuration with the lowest obtained
interwell potential barrier of the central molecule
(shown in Fig.~\ref{fig:rand3}) has no symmetry at all.
The orientational profile has three minima of different depth,
while the lowest interwell barrier is about 200 times lower
than the corresponding lowest barrier in the regularly
ordered lattice.

Also it should be noted that the molecules of
the regularly ordered lattice (namely, the molecules
with the $\phi_1$ orientation, see Fig.~\ref{fig:regular})
have the neighbourhood configuration with the highest
possible interwell potential barrier.
While minimizing the overall interaction energy,
this configuration minimizes also an interaction energy
at the minimum of the one molecule potential,
and deepens this minimum.

\section{Discussion}
The considered simple planar model recovers some of the features
of the fullerite lattice.
First of all, it predicts a multi-sublattice structure
for a system which would be arranged into a more
symmetric 1-sublattice manner in the absence of anisotropic
intermolecular interactions.

Then, the model involves lowering of orientation potential
relief of the molecule at the crystal surface.
This can be compared favourably to the absence
of H oriented molecules at the STM image
of the fullerite surface.\cite{WZWHLY01}
Futhermore, at a closer look this image shows a slight
difference in orientations of the fullerite molecules
belonging to the three sublattices which have the molecular 
$C_3$ rotation axes not perpendicular to the
surface. This difference is due to a competition of the
two-dimensional character of a surface (with probably another
subdivision into sublattices) with the bulk equilibrium
orientaitons of the molecules below the surface molecular layer.

Rather narrow character of the domain walls in the
considered high-symmetry system 
is rather natural for a system with only one kind 
of interaction involved.\footnote{%
For the case of ferromagnets the domain wall width is of 
the order of $a \sqrt{J/A}$, where $a$ is a lattice 
spacing, $J$ is an exchange, and $A$ is an anisotropy.
Since $A$ is a relativistic correction,
the ratio $J/A$ can be increased up to $10^6$.
But for the present case of one interaction 
this ratio is about~1.}
It agrees well with a very sharp character of a 
domain wall observed in a two-dimensional monolayer
of $C_{60}$ fullerene molecules.\cite{HYWLZYWCZ01}
This wall contains  also a very sharp kink, which is a kind
of essentially two-dimensional defect that can incorporate 
molecules with low orientational interwell barrierrs.

The sharp character of the observed kink implies
a possibility of existence of strongly localised
orientational defects also in the bulk of the three-dimensional
fullerite. Some of this strongly localised defects with
necessarily uncorrelated orientations of the neighbour
molecules orientations should involve molecules with
an orientational potential which is sufficiently shallow
to give a reasonable frequency of tunnelling transitions.
As to the rather high $C_{60}$ molecule mass, the recent
molecular dynamics simulations on the dislocation kink
tunnelling\cite{VSCJMR01}  in Ag show an efficiency of tunnelling
of complex heavy object under certain conditions.

The idea to explain negative thermal expansion of solids
with the double-well tunnelling statistics was suggested
by Freiman in 1983
for the case of solid methane.\cite{F83}
In the range of temperatures where the conventional phonon 
mechanism does not work the thermal expansion is established as 
a result of competition of two factors. The first factor is a 
lattice contraction due to the process of populating the tunnel states 
with an increase of temperature. Shrinking the distances between 
molecules increases the height of the orientational 
interwell barriers, what leads to a decrease of the tunnelling 
energy splitting, and as a result to a decrease of the 
system free energy. 
The contraction of the lattice is stabilised by an increase 
of an elastic part of the free energy for every fixed
value of crystal temperature.

Since the population of tunnel states has very strong exponential 
temperature dependence, the thermal expansion resulting from the 
competition of the two factors is always negative. 
At $T\approx 0$~K it is practically absent
(no molecules on excited tunnel levels).
With an increase of temperature the population of the excited state 
grows, therefore the lattice is contracted.
But at $T>\Delta$, where $\Delta$ is the tunnel state 
energy splitting, both the ground and the excited states 
become almost equally populated, so that the effect becomes 
much less pronounced.
This means that there should exist a maximum in the magnitude of 
the negative thermal expansion coefficient.

With a simple differentiation of expression (6) of 
Ref.~\cite{F83} one can obtain that this maximum takes place 
at the temperature $T_{\rm max}$ satisfying the equation 
\[
2T_{\rm max}/\Delta =
\tanh{(\Delta/2T_{\rm max} +
1/2\ln{(f_1/f_2)})},
\]
where $f_1$ and $f_2$ are the degeneracies of the ground
and the excited states, respectively.
It is easy to see that $T_{\rm max}<\Delta /2$ holds for any 
ratio $f_1/f_2$. 

Therefore the tunnelling energy splitting in fullerite can be estimated 
from the $T_{\rm max}$ position in Refs.~\cite{AEMSSU97,AGEMSUM01}
to be more than 8~K.
On the other hand, the positive thermal expansion of pure fullerite
at $T<2$~K implies a presence of processes other than 
two-well tunnelling (probably, the conventional 
phonon mechanism is still valid) at this low temperature.

The possibility to detect experimentally the negative 
contribution to thermal expansion due to the tunnelling 
objects depends strongly on the relative magnitude of the 
positive (conventional) and negative (tunelling in this case)
contributions. 
In the case of fullerite, the negative contribution is more pronounced,
but one still encounters a difficulty in determining
the tunnelling object.
The first hypothesis about a tunnelling of regular $C_{60}$
molecules between P and H orientations \cite{IL93} had a
drawback of high interwell potential barrier.
Further introduction of the idea of competition
of isotropic and anisotropic parts of intermolecular
interaction potential (though in orientational glass)
\cite{L99} has lead to a current understanding
(given in our previous paper \cite{LPK01}
and the present one) that the
tunnelling objects are to be found at the strongly
localised orientational defects of the fullerite structure.
The more detailed description of such defects could be
obtained with the help of more realistic three-dimensional
modelling of the $C_{60}$ crystal structure,
what should be a subject for future studies.

\section*{Acknowledgments}
We would like to thank Profs. A.S.Bakai, Yu.B.Gaididei, M.A.Ivanov
and V.G.Manzhelii, as well as Dr.A.N.Alexandrovskii for
valuable and critical discussions. The paper is partly supported by the
Program ``Investigation of Fundamental Problems and Properties
of the Matter on Micro- and Macrolevels" of 
the National Academy of Sciences of Ukraine,
by the INTAS Foundation under Grant INTAS 97-0368,
and also by the project N 2669 ``Structure
and plastisity of fullerite" of Science and 
Technology Center of Ukraine.


\small
\begin{thebibliography}{99}
\bibitem{L92}
V.M.~Loktev, \emph{Fiz.\ Nizkikh Temp.\ }\textbf{18}, 217 (1992)
[\emph{Low Temp.\ Phys.} {\bf 18}, 149 (1992)].
%old review
\bibitem{R94}
A.P.~Ramirez, \emph{Condens.\ Matter News} \textbf{3}, 6 (1994).
%old review
\bibitem{G97}
O.~Gunnarsson, \emph{Rev. Mod. Phys.} {\bf 69}, 575 (1997).
%review
\bibitem{KBK00}
H.~Kuzmany, B.~Burger, and J.~K\"{u}rty, in {\em Optical
and Electronic Properties of Fullerenes and
Fullerene-based Materials} / Ed. by J.~Shinar,
Z.V.~Vardeny, and Z.H.~Kaffafi (Marsel Dekker, New York,
2000).
%review

\bibitem{EMDN97}
V.B.~Efimov, L.P.~Mezhov-Deglin, and R.K.~Nikolaev,
\emph{ JETP Lett.} \textbf{65}, 687 (1997).
\bibitem{EMD99}
V.B.~Efimov, L.P.~Mezhov-Deglin,
\emph{ Physica B} \textbf{263-264}, 705 (1999).
%heat conduction

\bibitem{AEMSSU97}
A.N.~Aleksandrovskii, V.B.~Esel'son, V.G.~Manzhelii,
A.V.~Soldatov, B.~Sundquist, and B.G.~Udovidchenko,
\emph{Fiz.\ Nizkikh Temp.} \textbf{23}, 1256 (1997)
[\emph{Low Temp.\ Phys.} {\bf 23}, 943 (1997)];
\emph{ibid.} {\bf 26}, 100 (2000)
[{\bf 26}, 75 (2000)].
\bibitem{AGEMSUM01}
A.N.~Aleksandrovskii, V.G.~Gavrilko, V.B.~Esel'son,
V.G.~Manzhelii, B.~Sundqvist, B.G.~Udovidchenko,
and V.P.~Maletskiy,
\emph{Fiz.\ Nizkikh Temp.} \textbf{27}, 333 (2001)
[\emph{Low Temp.\ Phys.} {\bf 27}, 245 (2001)];
\emph{ibid.} {\bf 27}, 1401 (2001)
[{\bf 27}, 1033 (2001)].
%negative thermal expansion

\bibitem{OTP93}
J.P.~Olson, K.A.~Topp, and R.O.~Pohl,
\emph{Science} \textbf{259}, 1145 (1993).
%specific heat - experiment

\bibitem{YTSLM92}
R.C.~Yu, N.~Tea, M.B.~Salamon, D.~Lorents and R.~Malhotra,
\emph{Phys.\ Rev.\ Lett.} \textbf{23}, 2050 (1992).
\bibitem{SHY97}
S.~Savin, A.B.~Harris and T.~Yildirim,
\emph{Phys.\ Rev.\ B} \textbf{55}, 14 182 (1997).
%Energy estimates for single molecule orientations in fullerite

\bibitem{DIDP92}
M.~David, R.~Ibberson, T.~Dennis, and K.~Prassides,
\emph{Europhys.\ Lett.\ }\textbf{18}, 219 (1992).
% P and H configurations - first invented

\bibitem{TSB98}
S.P.~Tewari, P.~Silotia, and K.Bera,
\emph{ Solid State Comm.} \textbf{107}, 129 (1998).
%specific heat - tunnelling

\bibitem{IL93}
M.A.~Ivanov, V.M.~Loktev,
{\em Fiz.\ Nizk.\ Temp.\ }{\bf 19}, 618 (1993)
[{\em Low Temp.\ Phys.\ }{\bf 19}, 442 (1993)].
%Double well tunnelling for fullerite.

\bibitem{LPK01}
V.M.~Loktev, Yu.G.~Pogorelov, Ju.N.~Khalack,
{\em Fiz.\ Nizk.\ Temp.\ }{\bf 27}, 539 (2001)
[{\em Low Temp.\ Phys.\ }{\bf 27}, 397 (2001)].

\bibitem{LRM97}
P.~Launois, S.~Ravy, and R.~Moret,
\emph{Phys.\ Rev.\ B }{\bf 55}, 2651 (1997).
%intermolecular potential modelling

\bibitem{L_af}
V.M.~Loktev,
{\em Fiz.\ Nizk.\ Temp.\ }{\bf 5}, 295 (1979)
[{\em Sow.\ J.\ Low Temp.\ Phys.\ }{\bf 5}, 142 (1979)].
%3-sublattice antiferromagnet structure

\bibitem{WZWHLY01}
H.~Wang, C.~Zeng, B.~Wang, J.G.~Hou, Q.~Li, J.~Yang,
{\em Phys.\ Rev.\ B} {\bf 63}, 085417 (2001).
%STM of fullerite surface

\bibitem{TAVLM92}
G.~Van Tendeloo, S.~Amelinckx, M.A.~Verheijen,
P.H.M.~van Loosdrecht, G.~Meijer,
{\em Phys.\ Rev.\ Lett.} {\bf 69}, 1065 (1992).
%Electron-diffraction 8 sublattice C$_{60}$ superstructure

\bibitem{GPMSWM92}
E.J.J.~Groenen, O.G.~Poluektov, M.~Matsushita,
J.~Schmidt, J.H.~van der Waals, G.~Meijer,
{\em Chem.\ Phys.\ Lett.} {\bf 197}, 314 (1992).
%Triplet excitation 8 sublattice C$_{60}$ superstructure at 1.2 K

\bibitem{LPHLT01}
C.~Laforge, D.~Passerone, A.B.~Harris, P.~Lambin,
E.~Tosatti,
{\em Phys.\ Rev.\ Lett.} {\bf 87}, 085503 (2001).
%C60 surface modelling: two-stage disorder

\bibitem{HYWLZYWCZ01}
J.G.~Hou, J.~Yang, H.~Wang, Q.~Li, C.~Zeng,
l.~Yuan, B.~Wang, D.M.~Chen, Q.~Zhu,
{\em Nature} {\bf 409}, 304 (2001).

\bibitem{VSCJMR01}
T.~Vegge, J.P.~Sethna, S.A.~Cheong, K.W.~Jacobsen,
C.R.~Myers, D.C.~Ralph,
{\em Phys.\ Rev.\ Lett.} {\bf 86}, 1546 (2001).

\bibitem{F83}
Yu.A.~Freiman,
{\em Fiz.\ Nizk.\ Temp.\ }{\bf 9}, 657 (1983).

\bibitem{L99}
V.M.~Loktev, \emph{Fiz.\ Nizkikh Temp.\ }\textbf{25}, 1099 (1999)
[\emph{Low Temp.\ Phys.} {\bf 25}, 823 (1999)].

\end{thebibliography}
\end{document}